\documentclass[
preprint,tightenlines,
nofootinbib,superscriptaddress,amsmath,amssymb,aps,prd]{revtex4-2}

\usepackage[utf8]{inputenc}
\usepackage{graphicx}
\usepackage{hyperref}
\usepackage{url}
\usepackage{color}
\usepackage{isotope}

\def\be{\begin{equation}}
\def\ee{\end{equation}}
\def\bea{\begin{eqnarray}}
\def\eea{\end{eqnarray}}

\newcommand{\cevns}{CE$\nu$NS}
\newcommand{\xst}{$X_{17}$}

\begin{document}

\title{\textbf{\Large Glimpses of the \xst\ from \\[0.25cm]  coherent elastic neutrino nucleus scattering }}

\author{J.~Rathsman}
\email{Corresponding author: johan.rathsman@fysik.lu.se}
\affiliation{Department of Physics, Lund University,  221 00 Lund, Sweden.}

\author{J.~Cederk{\"a}ll}
\email{joakim.cederkall@fysik.lu.se}
\affiliation{Department of Physics, Lund University, 221 00 Lund, Sweden.}

\author{Y.~Hi\c{c}y{\i}lmaz}
\email{yasarhicyilmaz@balikesir.edu.tr}
\email{yasar.hicyilmaz@physics.uu.se}~
\email{y.hicyilmaz@soton.ac.uk}
\affiliation{Department of Physics, Bal{\i}kesir University,   TR10145, Bal{\i}kesir, Turkey.}
 \affiliation{Department of Physics and Astronomy, Uppsala University, 751 20, Uppsala, Sweden.} 
 \affiliation{School of Physics and Astronomy, University of Southampton, Highfield, Southampton SO17 1BJ, United Kingdom.} 

\author{E.~Lytken}
\email{else.lytken@fysik.lu.se}
\affiliation{Department of Physics, Lund University,  221 00 Lund, Sweden.}

\author{S.~Moretti}
\email{stefano.moretti@physics.uu.se}
\email{s.moretti@soton.ac.uk}
\affiliation{Department of Physics and Astronomy, Uppsala University,  751 20, Uppsala, Sweden.}
\affiliation{School of Physics and Astronomy, University of Southampton, Highfield, Southampton SO17 1BJ, United Kingdom.}

\begin{abstract}
We show that the process of Coherent Elastic neutrino ($\nu$) Nucleus Scattering (\cevns) at nuclear reactor experiments has 
significant sensitivity to the so-called \xst\ particle, 
which has been invoked to explain the ATOMKI anomaly, wherein electron-positron pairs emerging from a nuclear transition of excited $^8$Be, $^4$He and $^{12}$C nuclei are studied. Such a new state has  potentially been identified as a spin-1 object, with axial-vector couplings and  a mass around 16.7 MeV, hence, in the kinematic range accessible by the aforementioned experimental settings. Specifically, we fit CONUS+ and Dresden-II data and show that a robust statistical analysis renders these more compatible with the \xst\ hypothesis, in turn interfering with the Standard Model (SM), than with that of the latter alone. The same stays true when also adding  COHERENT data from $\pi^+$ decays at rest, singling out two regions of preferred couplings of the \xst\ to electron and muon neutrinos as well as nuclei.  
\end{abstract}    
\maketitle

\clearpage
\section{Introduction}

Coherent Elastic neutrino ($\nu$) Nucleus Scattering (\cevns) involves the scattering of a neutrino off an atomic nucleus as a whole, and has promise to be a key process for probing both Standard Model (SM) and Beyond the SM (BSM) interactions. In this paper we analyze data from the first two experiments that have published reactor neutrino results using \cevns, and combine them with \cevns\ data from the COHERENT experiment at Oak Ridge National Laboratory (ORNL) as well as neutrino data from the ICEcube experiment, to investigate the potential existence of a Z' boson.

The \cevns\ process was first suggested by Freedman in 1973~\cite{Freedman:1973yd}, but the small nuclear recoil energies involved meant that it could not be experimentally verified until 2017, when the COHERENT collaboration~\cite{COHERENT:2017ipa} confirmed its existence using secondary neutrinos produced by the Spallation Neutron Source (SNS) at ORNL. The first measurement of \cevns\ using reactor anti-neutrinos was carried out at the Dresden-II reactor~\cite{Colaresi:2022obx}, Morris, Il, USA, and subsequently confirmed by the CONUS+ collaboration~\cite{Ackermann:2025obx} in 2025 at the Leibstadt nuclear power plant, Switzerland. Reactors provide a constant high-flux source of low-energy (a few MeV) electron anti-neutrinos, whereas the spallation sources produce higher-energy (tens of MeV), pulsed  and multi-flavour fluxes.

In addition to precision neutrino measurements, neutrinos from reactors can also be used to search for new low-mass gauge bosons, e.g., emerging from $U(1)^\prime$ theories. Among such $U(1)'$ frameworks, we focus here on a case that embeds a state that can be identified as the \xst\ particle invoked to explain the ATOMKI anomaly \cite{Krasznahorkay:2015iga,Sas:2022pgm,Krasznahorkay:2017gwn,Krasznahorkay:2017bwh,Krasznahorkay:2017qfd,Krasznahorkay:2018snd,
Krasznahorkay:2019lyl,Krasznahorkay:2021joi,Krasznahorkay:2022pxs,Krasznahorky:2024adr}.
The ATOMKI experiment, at the Hungarian Institute for Nuclear Research in Debrecen \cite{Gulyas:2015mia}, has over the past few years studied the properties of $e^+e^-$ pairs generated during nuclear transitions in $^8{\rm Be}$, $^4{\rm He}$ and $^{12}{\rm C}$ nuclei 
\cite{Krasznahorkay:2015iga,Sas:2022pgm,Krasznahorkay:2017gwn,Krasznahorkay:2017bwh,Krasznahorkay:2017qfd,Krasznahorkay:2018snd,
Krasznahorkay:2019lyl,Krasznahorkay:2021joi,Krasznahorkay:2022pxs,Krasznahorky:2024adr}.
An excess in the invariant mass and opening angle has consistently been observed, at more than 5$\sigma$ level when combined across the studied cases, suggesting the possibility of a new particle with a mass of approximately 17~MeV.  
Additional evidence of the \xst\ has emerged from an independent experiment at 
VNU University of Science in Hanoi, which reported replication of the ATOMKI anomaly in early 2024, from a measurement of the angular distribution for the $^8$Be case\cite{Anh:2024req}. 

The PADME experiment 
at the Laboratori Nazionali di Frascati (LNF) has also shown sensitivity to this mass range and has recently posted a result in Ref.~\cite{PADME:2025dla},
stating a $2\sigma$  excess at the mass indicated by the ATOMKI experiment. In contrast, the MEG-II experiment at the Paul Scherrer Institut (PSI), Villingen, Switzerland, appears to disfavour the \xst\ hypothesis and has set an upper limit on the $e^+e^-$ decay rate of such a possible new particle 
\cite{MEGII:2024urz}. Other experiments also have sensitivity to a possible \xst\ state, including the NA62 and nTOF experiments at CERN, the Mu3e experiment at PSI, the PRad, DarkLight and HPS experiment at JLab and the New JEDI project at GANIL. Still, none of the exeriments already operating has reported conclusive arguments in either direction \cite{Alves:2023ree}. 

Interpretations of the \xst\ have been numerous in the literature. We focus, in this paper, on the assumption that the \xst\ is a spin-1 boson, which could well, but not necessarily, be the carrier of a postulated fifth force, potentially connected with a dark sector, itself containing a Dark Matter (DM) candidate. In Ref.~\cite{Feng:2016jff} it is noted that, in the case of a pure vector coupling, the new boson should be protophobic, as implied primarily by the NA48/2 and NA64  data~\cite{NA482:2015wmo,NA64:2019auh} (see \cite{Feng:2016ysn,Ellwanger:2016wfe,Feng:2020mbt,Nomura:2020kcw,Seto:2020jal,Kozaczuk:2016nma,DelleRose:2018pgm,DelleRose:2019hnc,DiLuzio:2025ojt} for  alternative interpretations).
Further phenomenological studies~\cite{Zhang:2020ukq,Barducci:2022lqd,Denton:2023gat} have concluded that, due to absence of bremsstrahlung from the \xst, and the non-observation of deviations from the SM in neutrino scattering experiments, theoretical frameworks including a pure vector  mediator  are less likely,  while an axial-vector state appears as the best candidate to comply with the mentioned restrictions.

In addition, several analyses have been conducted to investigate the effects of a light \xst\ particle  on the anomalous magnetic moment of the electron ($a_e$) and muon ($a_\mu$), including flavour observables (such as $R_{K^{(*)}}$)~\cite{Nomura:2020kcw,Seto:2020jal,DelleRose:2017xil,DelleRose:2019ukt,Barman:2021yaz,Bodas:2021fsy,Fayet:2020bmb,Hati:2020fzp}. Presently, these data sets all indicate consistency with the existence of an \xst\ particle, and the possibility of detecting it in the future. For other attempts to interpret the ATOMKI results in terms of background nuclear physics or QCD effects we refer to~\cite{Zhang:2017zap,Koch:2020ouk,Chen:2020arr,Aleksejevs:2021zjw,Kubarovsky:2022zxm,Hayes:2021hin,Viviani:2021stx}. 

A minimal theoretical approach embedding the \xst\ is given by a family-dependent $U(1)$ extension of the SM, where symmetry breaking introduces a new light vector boson, a $Z'$, which can be interpreted as the \xst, as first introduced in Ref.~\cite{DelleRose:2018eic}. This approach allows for axial-vector couplings, depending on the gauge quantum numbers involved. Furthermore, in this framework, the Yukawa interactions are suitably modified by higher-dimensional operators~\cite{Pulice:2019xel}. This construction is what we will pursue in the present paper.  We note that, in a recent study~\cite{Fieg:2026zkg}, an \xst\ with both vector and axial vector couplings has been put forward, although that particular realisation was found to be in tension with constraints from atomic parity violation. Similarly a model with chiral \xst\ couplings and two Higgs doublets has been proposed in~\cite{Batra:2026tzz}.

One should notice that the proposed scenario does not prevent the \xst\ from coupling to neutrinos, which means it can also lead to Non-Standard Interactions (NSIs) affecting neutrino flavour ratios in matter \cite{Proceedings:2019qno}. Therefore, the experimental constraints on NSIs from neutrino oscillations can be applied to restrict the family dependent, non-universal, couplings of the new boson to SM fermions, as was recently done in Ref.~\cite{Enberg:2024ofo} using data from the TEXONO~\cite{TEXONO:2009knm,Bilmis:2015lja} and IceCube experiments~\cite{IceCubeCollaboration:2021euf}. The possibility of the \xst\ coupling to neutrinos, has also prompted us to study, in the spirit of Ref.~\cite{Chattaraj:2025rtj}, the possibilities to discover this object using \cevns\ at the European Spallation Source (ESS) in Lund~\cite{Cederkall:2025bka}, building on the  work done in Ref.~\cite{Procs}. Similar results, using \cevns\, may also be obtained from the COHERENT experiment \cite{COHERENT:2020iec,COHERENT:2020ybo,COHERENT:2021xmm,COHERENT:2023aln,COHERENT:2024axu}\footnote{See https://indico.cern.ch/event/1439855/contributions/6461655/.}.

In light of the strong sensitivity to the \xst\ displayed by CE$\nu$NS experiments at ESS, or other spallation sources, we revisit here the possibilities  offered by nuclear reactor experiments such as
Dresden-II and CONUS+ as well as others (TEXONO~\cite{TEXONO:2024vfk}, $\nu$GEN~\cite{nGeN:2022uje}, NUCLEUS~\cite{NUCLEUS:2019igx}, NEON~\cite{NEON:2022hbk}, CONNIE~\cite{CONNIE:2021ggh}) which are in the planning or have not yet made any observations.
\cevns\ measurements from Dresden-II and CONUS+  have also been included in the BSM analyses of Refs.~\cite{AtzoriCorona:2022qrf,Chattaraj:2025fvx, DeRomeri:2025csu,AtzoriCorona:2025ygn,AtzoriCorona:2025xgj}, although, as pointed out in~\cite{Li:2025pfw}, there is some tension between the data from Dresden-II and CONUS+ when analysed in the SM. Very recently, also the impact of light mediators contributing through \cevns\ to DM direct detection experiments, using data from XENON~\cite{XENON:2024ijk} and PandaX~\cite{PandaX:2024muv}, has been been analysed~\cite{AtzoriCorona:2025gyz,DeRomeri:2025nkx}. 
On a related note, reactor data from \cevns\ has also been used to improve the determination of the neutrino mass ordering~\cite{Denton:2022nol}.

More generally,  \cevns\ can also be used to study phenomena within the SM, including the neutron skin effect in nuclei and the neutrino charge radius (see~\cite{Cadeddu:2019eta} for an early example), as well as  BSM physics (especially when its mass states are in the sub-GeV  range)~\cite{Lindner:2016wff}. In particular, light vector bosons such as the \xst\ can contribute to the cross-section both through constructive and destructive interference with the $\gamma$ and $Z$ bosons of the SM. Until our recent study~\cite{Cederkall:2025bka,Procs}, only flavour universal models  with a light $Z^\prime$ in this mass range, e.g., the $B-L$ one, where the couplings to electron and muon neutrinos are the same, have been considered~\cite{Abdullah:2018ykz,Flores:2020lji,Cadeddu:2020nbr,Banerjee:2021laz,AtzoriCorona:2022moj,Coloma:2022avw}. For a brief review of the physics of \cevns\ we refer to~\cite{Cadeddu:2023tkp}.

The layout of our paper is  as follows. In the next section we introduce our BSM framework, including a description of its implementation using different numerical tools. We proceed to describe the theory of \cevns\ for nuclear reactors in Sect.~\ref{Sect3}, where we also discuss the experimental setups that presently show sensitivity to the \xst. This includes interpreting their data in combination with those from the COHERENT experiment. The technical details of our treatment of the COHERENT detector performance for carrying out the data combination discussed above, are left for an Appendix. We conclude in Sect.~\ref{Sect4}.

\section{The theoretical framework embedding the \xst\ }
\label{Sect2}

In the following we give a brief description of the minimal extension of the SM with a generic $U(1)'$  gauge group studied in this work. For more details of the model we refer to~\cite{Cederkall:2025bka}. The kinetic terms in the  Lagrangian that depend on the hypercharge group of the SM,  $U(1)_Y$, and the additional gauge structure, $U(1)'$, are  given by
\begin{equation}
\label{eq:KineticL}
\mathcal{L}_\mathrm{Kin}^{U(1)'} = - \frac{1}{4} \hat F_{\mu\nu} \hat F^{\mu\nu}  - \frac{1}{4} \hat F'_{\mu\nu} \hat F^{'\mu\nu} - \frac{\eta}{2} \hat F'_{\mu\nu} \hat F^{\mu\nu},
\end{equation}
where the field strengths $  \hat F_{\mu\nu} $  and $ \hat  F'_{\mu\nu} $ correspond to the gauge fields $  \hat B_\mu $ and $ \hat  B'_\mu$ of $U(1)_Y$ and $U(1)'$, respectively, while the parameter $\eta$ quantifies the kinetic mixing between these Abelian symmetries. The gauge covariant derivative is written as
\begin{equation}
\label{CovDer}
{\cal D}_\mu = \partial_\mu + \dots + i g_1 Y B_\mu + i (\tilde{g} Y + g' Q') B'_\mu, 
\end{equation}
where $Y$ represents the hypercharge with coupling $g_1$ while $Q'$ denotes the $U(1)'$ charge with associated coupling $g'=\hat g'/\sqrt{1-\eta^2}$, where $\hat g'$ is the original $U(1)'$ coupling before the transformation of the gauge fields for diagonalisation, while $\tilde{g}= -\eta g_1/\sqrt{1-\eta^2}$ characterises the strength of the gauge mixing between $ U(1)_{Y} $ and $ U(1)' $.

In the model, a new SM singlet complex scalar $ \chi $ breaks the $U(1)'$ symmetry spontaneously
through  a Vacuum Expectation Value (VEV)  $\langle\chi\rangle =  v'/\sqrt{2}$. A new massive vector boson is generated by such a breaking of $U(1)'$. The neutral gauge boson mass eigenstates in the model are obtained through the rotation:
\bea
\left( \begin{array}{c} B^\mu \\ W_3^\mu \\ B'^\mu \end{array} \right) = \left( \begin{array}{ccc} 
	\cos \theta_W & - \sin \theta_W \cos \theta' & \sin \theta_W \sin \theta' \\
	\sin \theta_W & \cos \theta_W \cos \theta' & - \cos \theta_W \sin \theta' \\
	0 & \sin \theta'  & \cos \theta'  
\end{array} \right)
\left( \begin{array}{c} A^\mu \\ Z^\mu \\ Z'^\mu \end{array} \right),
\eea
where $\theta_W$ is the Weinberg angle and $\theta'$ is the $Z-Z'$ mixing angle \cite{Coriano:2015sea}:
\bea \label{ThetaPrime}
\tan 2 \theta' = \frac{2 g_H g_Z }{ g_H^2 + ( 2 Q'_\chi g' \, v'/v )^2 - g_Z^2} \,,
\eea
with $Q'_\chi$ being the $U(1)'$  charge of the $\chi$-field, $ g_H = \tilde{g} + 2 g' Q'_H $ and $g_Z=\sqrt{g_2^2 + g_1^2} $ is the Electro-Weak (EW) coupling. 

The mixing angle $\theta'$ is constrained by the LEP precision measurements to be such that $|\theta'| < 10^{-3}$ \cite{DELPHI:1994ufk,Erler:2009jh}. In the  small $\theta'$ approximation, the $Z$ and $Z'$ masses can be written as 
\bea \label{ZZpMassesSmallTh}
m_Z^2 \simeq  \frac{v^2}{4} g_Z^2
\,, \qquad 
m_{Z'}^2 \simeq ( Q'_\chi g' \, v')^2,
\eea
while the mixing angle simplifies to
\bea \label{ThetaPrimeMasses}
\theta' \simeq -\frac{g_H}{g_Z}  \,,
\eea
where $g_H^2 \ll g_Z^2$ and $(g' v')^2 \ll (g_Z v)^2$  have also been assumed.
For $g' \sim {\cal O}(10^{-4} - 10^{-5})$, $M_{Z'}$ becomes light (${\cal O}(10)$ MeV) when $ v' \approx {\cal O}(100 -1000) $ GeV,  then potentially offering a candidate to address the ATOMKI anomaly.

The Lagrangian for the $  Z' $ interaction with SM fermions is given by
\begin{eqnarray}
\label{eq:NeuCurLag}
\mathcal{L}^\mathrm{Z'} &=& \bar q \gamma^\mu \left( C^{qq'}_{L} P_L + C^{qq'}_{R} P_R \right) q' Z'_\mu + \bar \nu_l \gamma^\mu \left( C^{ll'}_{L} P_L \right) \nu_{l'} Z'_\mu \nonumber \\ 
&+&\bar l \gamma^\mu \left( C^{ll'}_{L} P_L + C^{ll'}_{R} P_R \right) l' Z'_\mu,
\end{eqnarray}
where $ q^{(\prime)}$,  $l^{(\prime)}$  and $  \nu_{l^{(\prime)} }$ refer to up-type/down-type quarks, charged leptons and their neutrinos, while $ C^{XX}_{L} $ and $ C^{XX}_{R} $ are Left ($L$) and Right ($R$) handed couplings, respectively, with $P_L$ and $P_R$ the corresponding projection operators $\frac{1\mp\gamma^5}{2}$. In this model, we assume a flavour-conserving scenario for the quark and lepton sectors giving diagonal couplings according to
\begin{eqnarray}
	C^{ff}_{L} \!&\!=\!&\!  - g_Z \sin \theta' \left( T^3_f - \sin^2 \theta _W Q_f \right) \!+\! ( \tilde g Y_{f, L} \!+\! g' Q'_{f, L})  \cos \theta'\!, \label{CffL}  \\
	C^{ff}_{R} &=&  g_Z \sin^2 \theta_W \sin\theta' Q_f + ( \tilde g Y_{f, R} + g' Q'_{f, R}) \, \cos \theta ', \label{CffR}
	\end{eqnarray}
where $T_f ^3$ and $Q_f$ are the fermion's weak isospin and electric charge, respectively, whereas $Y_{f,L/R}$ and $Q'_{f,L/R}$ denote  the $U(1)_Y$ and $U(1)'$ charges. 

As for the vector and axial couplings, these are defined as
\bea \label{VAcouplings}
C_{f, V} = \frac{C^{ff}_{R} + C^{ff}_{L}}{2} \,, \quad C_{f, A} = \frac{C^{ff}_{R} - C^{ff}_{L}}{2} .
\eea
 In the limit of small gauge coupling and mixing, $ g^\prime, \tilde{g} \ll 1$, the vector and axial couplings are found as~\cite{DelleRose:2018pgm}
\begin{align}
\label{eq:couplings}
&C_{f, V} \simeq \tilde{g} \cos^2\theta _W Q_f  + g^\prime [Q'_H (T^3_f  - 2 \sin^2\theta _W Q_f ) + (Q'_{f, R}+Q'_{f, L})/2]\, , \nonumber \\
&C_{f, A} \simeq g^\prime [-Q'_H T^3_f + (Q'_{f, R}-Q'_{f, L})/2]\, .
\end{align}

Concerning the Yukawa sector of the SM Lagrangian, this reads as:
\be 
- \mathcal{L}_{\rm Yuk.}^{\rm SM} = Y_u \bar{Q} \tilde{H} u_R + Y_d \bar{Q} H d_R + Y_e \bar{L} H e_R \, .
\label{Yukawa}
\ee

For the $U(1)'$ charges, satisfying the gauge invariance conditions in such a Yukawa sector, axial-vector $Z'$ couplings for quarks and charged leptons vanish. The experimental results put very stringent constraints on a pure vector \xst. Instead, our model introduces flavour-dependent $U(1)'$ charges to generate crucial axial-vector couplings with nucleons, avoiding experimental constraints \cite{Feng:2016jff,Feng:2016ysn,NA482:2015wmo}. The model provides this through a mechanism where only third-generation fermions obtain masses via SM-like Yukawa terms, while the first- and second-generation masses arise from higher-dimensional operators:
\bea
\label{eq:powers}
- \mathcal{L}_{\rm Yukawa} &=& 
\Gamma^{u}_i \left( \frac{\chi^*
}{M} \right)^{n_{i}} \overline{Q}_{L,i}\tilde{H}u_{R,i} 
+ \Gamma^{d}_i\left( \frac{\chi}{M} \right)^{l_{i}} \overline{Q}_{L,i} H d_{R,i} \nonumber \\
&+&\Gamma^{e}_i \left(\frac{\chi}{M} \right)^{m_{i}} \overline{L}_{i} H e_{R,i}+ h.c., \quad i=1,2 \, ,
\eea
where $M$ sets the non-renormalisable scale and we assume that the couplings are flavour diagonal in order to avoid tree level flavour changing neutral currents. The requirement of gauge invariance under the $U(1)'$ symmetry then gives
\begin{align}
\label{eq:gauge_invariance_gen12}
& -n_{1,2} Q'_{\chi}-Q'_{Q_{1,2}} -Q'_H +Q'_{u_{1,2}}  = 0, \nonumber \\
& l_{1,2} Q'_{\chi}-Q'_{Q_{1,2}} + Q'_H +Q'_{d_{1,2}}  = 0,  \nonumber \\
&m_{1,2} Q'_{\chi} -Q'_{L_{1,2}} + Q'_H +Q'_{e_{1,2}}   = 0,
\end{align}
which will generate  non-zero axial couplings when $n_i$, $m_i$ or $l_i\ne0$. In the following we assume flavour-universality of the first two generations of $U(1)'$ quark charges ($Q'_{Q_1} = Q'_{Q_2}$ etc.), 
whereas the lepton charges are fully non-universal. As a consequence we  have $ n_{1} =n_{2} $ and $ l_{1} =l_{2} $. The $U(1)'$ charges must also satisfy the anomaly cancellation conditions for the SM fermions as well as additional  $R$-handed neutrinos: 
\begin{align}
\label{eq:anomaly}
& \sum_{i=1}^{3} (2 Q'_{Q_i} - Q'_{u_i} - Q'_{d_i}) = 0 \,,  \\
&  \sum_{i=1}^{3} \, ( 3 Q'_{Q_i} +  Q'_{L_i})  = 0 \,,  \\
& \sum_{i=1}^{3} \left( \frac{Q'_{Q_i}}{6} - \frac{4}{3} Q'_{u_i} - \frac{Q'_{d_i}}{3}  +   \frac{Q'_{{L_i}}}{2} - Q'_{e_i}\right) = 0 \,, 
\label{eq:anomaly3} \\
& \sum_{i=1}^{3} \left( Q_{Q_i}^{\prime 2} - 2 Q_{u_i}^{\prime 2}  + Q_{d_i}^{\prime 2}       - Q_{{L_i}}^{\prime 2}  + Q_{e_i}^{\prime 2}  \right) = 0 \,, \label{eq:anomaly4} \\
& \sum_{i=1}^{3} \left( 6 Q_{Q_i}^{\prime 3}  - 3 Q_{u_i}^{\prime 3}  - 3 Q_{d_i}^{\prime 3}  +  2 Q_{{L_i}}^{\prime 3}  - Q_{e_i}^{\prime 3} \right)  + \sum_{i=1}^{3} Q_{\nu _i}^{\prime 3}    = 0 \,, \label{eq:anomaly5} \\
& \sum_{i=1}^{3} \left( 6 Q'_{Q_i} - 3 Q'_{u_i} - 3 Q'_{d_i}  + 2 Q'_{{L_i}} - Q'_{e_i} \right) + \sum_{i=1}^{3} Q'_{\nu _i}    = 0 .
\label{eq:anomaly6}
\end{align}
Further information about the general charge solutions of our model can be found in Ref.~\cite{Cederkall:2025bka}.

As already mentioned, in addition to the SM Higgs sector, our model includes the new complex scalar $ \chi $ whose VEV is responsible for spontaneous symmetry breaking of the $U(1)'$ symmetry. The scalar potential of the model is given by
\begin{equation}
\label{eq:HM}
V(H,\chi) =  -\mu^2|H|^2 +  \lambda |H|^4 -\mu_\chi^2 |\chi|^2 + \lambda_\chi |\chi|^4 +\kappa  |\chi|^2|H|^2,  
\end{equation}
where $H$ is the SM Higgs doublet and $\kappa$ is the mixing parameter coupling the SM Higgs field to the new scalar $ \chi $. As mentioned, this model features two physical Higgs states with two non-vanishing VEVs, $v$ and $v'$, which are defined as follows:
\bea
\label{eq:vev}
\langle H \rangle = \frac{1}{\sqrt{2}} \left( \begin{tabular}{c} 0 \\ $v$ \end{tabular} \right) \,, \qquad \langle\chi\rangle = \frac{v'}{\sqrt{2}}\,.
\eea
The physical CP-even Higgs mass eigenstates  ($h_1$ and $h_2$) emerge from the interaction states ($H$,$ \chi $) through an orthogonal rotation:
\bea
\left( \begin{array}{c} h_1 \\ h_2 \end{array} \right) = \left( \begin{array}{cc} \cos \theta & - \sin \theta \\  \sin \theta & \cos \theta \end{array} \right)  \left( \begin{array}{c} H  \\ \chi \end{array} \right),
\eea 
with the mixing angle $\theta$ constrained to  $- \pi/2 < \theta < \pi/2$. The corresponding mass eigenvalues are given by:
\bea
m_{h_{1,2}}^2 = \lambda v^2 + \lambda_\chi  v'^2 \mp \sqrt{\left( \lambda v^2 - \lambda_\chi  v'^2\right)^2 + \left( \kappa v v' \right)^2}.
\eea
The mixing angle can be determined from parameters of the scalar potential via 
\bea
\tan 2 \theta = \frac{\kappa v v'}{\lambda v^2 - \lambda_\chi  v'^2}.
\label{theta}
\eea
Here, we assume $h_2$ as the SM-like Higgs boson while the exotic boson $h_1$ is lighter and dominantly a singlet-like Higgs state, which has been considered in~\cite{Hicyilmaz:2022owb} as a mediator for possible $Z^\prime$ signatures.

The above summarises the \xst\ model that we are considering. The free parameters of the model are given in Tab.~\ref{tab_paramSP}, alongside their scanned ranges. To analyse the model we have implemented it using different numerical tools as follows. We have used the SPheno~\cite{Porod:2003um,Porod:2011nf,Braathen:2017izn} and SARAH 4.14.3~\cite{Staub:2013tta,Staub:2015kfa} codes for the parameter space investigation of the model. Scanning of the parameter space was performed using linear samplings, within the ranges specified in Tab. \ref{tab_paramSP}. Moreover, we implement experimental constraints to our solutions. We first require the SM Higgs boson mass to be within $3$~GeV of its observed value of 125 GeV and its Branching Ratios (BRs), dominated by BR$(h_2 \to b\bar{b})$ and BR$(h_2 \to {\rm invisible}~[{\it i.e.}~h_1 h_1,Z' Z', Z Z'])$  to be within the experimental limits~\cite{ParticleDataGroup:2024cfk,ATLAS:2023tkt}. Furthermore, we apply the constraints on the BRs of rare $B$-decays, specifically $ {\rm BR}(B \rightarrow X_{s} \gamma) $ and $ {\rm BR}(B_u\rightarrow\tau \nu_{\tau}) $ \cite{HFLAV:2012imy,HFLAV:2010pgm}. We have also bounded the value of the $ Z-Z' $ mixing parameter $\theta'$ (see Eq.~(\ref{ThetaPrime})) to be less than a few times $ 10^{-3} $ as a result of EW precision observables~\cite{DELPHI:1994ufk,Erler:2009jh}.

\begin{table}[t!]
	\centering
	\begin{tabular}{c|c||c|c}
		\hline
		Parameter  & Scanned range & Parameter      & Scanned range \\
		\hline
		$g'$ & $[10^{-5}, 2\times 10^{-4}]$      & $\lambda$ & $[0.124, 0.127]$ \\
		$\tilde{g}$        & $[-7\times10^{-4}, 7\times10^{-4}]$ & ${\lambda}_{\chi}$ & $[0, 0.015]$ \\
		$v'$ & $[0.1, 1.5]$ TeV  &  $\kappa$  & $[0, 0.01]$ \\
		$   Q'_{d_3},Q'_{H}$ & $[-1, 1]$ &$Q'_{\nu_1} $& $0$ \\
        $  |Q'_{\chi}|$ & $[0.1, 2]$ & $Q'_{e_3}, Q'_{e_2}$ & $[-2, 2]$ \\
        $  m_{1} $ & $0$ & $m_{2}$ &  $2$ \\
		\hline
	\end{tabular}
	\caption{Scanned parameter space of the theoretical model.    }
	\label{tab_paramSP}
\end{table}
In the next step of the analysis, we constrain the parameter space according to the current experimental bounds on the anomalous magnetic moment result for
$ (g-2)_{\mu} $ \cite{Aliberti:2025beg,Muong-2:2025xyk}, the related $Z'$ couplings satisfying the ATOMKI anomaly~\cite{Barducci:2022lqd}, the electron beam dump experiment NA64~\cite{NA64:2019auh} and $e^+e^-$ collider experiment KLOE~\cite{Anastasi:2015qla} which probe the $Z'$ couplings to electrons,  $ C^2_{e,V}+ C^2_{e,A} $. We have also applied additional experimental limits on the neutrino scattering from the TEXONO experiment~\cite{TEXONO:2009knm}, parity-violating Moller scattering~\cite{SLACE158:2005uay,Kahn:2016vjr} and atomic parity violation~\cite{Porsev:2009pr,Arcadi:2019uif}.  
In addition to these constraints, NA48/2~\cite{NA482:2015wmo}, IceCube NSI~\cite{IceCubeCollaboration:2021euf} constraints on $Z'$ couplings and NA64 bounds for the invisible $Z'$ decays~\cite{NA64:2023wbi} have been applied. The full experimental constraints are summarised in Tab.~\ref{tab:constraints}. 
\begin{table}[t!]
\centering
\begin{tabular}{c|c|c}
\hline
Observable & Constraint & Reference(s) \\		
\hline
 $m_{h_2}$ & 122 GeV -- 128 GeV  & ~ \\
 BR$(h_2 \to b\bar{b})$ & $>0.4$&  \cite{ParticleDataGroup:2024cfk} \\
 BR$(h_2 \to {\rm invisible})$ & $<0.17$ & \cite{ATLAS:2023tkt} \\
 ${\rm BR}(B \rightarrow X_{s} \gamma)$ &
$2.99 \times 10^{-4}$ --
$3.87 \times 10^{-4}$ &
\cite{HFLAV:2012imy} \\
 $\frac{{\rm BR}(B_u\rightarrow\tau \nu_{\tau})}{{\rm BR}(B_u\rightarrow \tau \nu_{\tau})_{\rm SM}}$ &
0.15  --
2.41 &
\cite{HFLAV:2010pgm} \\
 $\Delta a_{\mu}$ &
$(3.8 \pm 6.3)\times 10^{-9}$ &
\cite{Aliberti:2025beg,Muong-2:2025xyk} \\
 $ \sqrt{C^2_{e,V}+ C^2_{e,A}} $ &
$\geq 2 \times 10^{-4} / \sqrt{{\rm BR}(Z' \to e^+ e^-)}$ &
\cite{NA64:2019auh} \\
 $ \sqrt{C^2_{e,V}+ C^2_{e,A}} $ &
$\leq 6 \times 10^{-4} / \sqrt{{\rm BR}(Z' \to e^+ e^-)}$ &
\cite{Anastasi:2015qla} \\
 $\epsilon_{ee}^{\oplus }-\epsilon_{\mu\mu}^{\oplus }$ &
$[-2.26, -1.27] \cup [-0.74, 0.32]$ &
\cite{IceCubeCollaboration:2021euf} \\
 $\epsilon_{\tau \tau}^{\oplus }-\epsilon_{\mu\mu}^{\oplus } $ &
$[-0.041,0.042]$ &
\cite{IceCubeCollaboration:2021euf} \\
 $ \sqrt{C_{e,V} \times C_{\nu_e,V}} $ &
$\leq 9 \times 10^{-5} $ &
\cite{TEXONO:2009knm} \\
 $ |C_{e,V} \times C_{e,A}| $ &
$\leq 10^{-8}  $ &
\cite{SLACE158:2005uay,Kahn:2016vjr} \\
 $ |C_{e,A}| \left| \frac{188}{399}  C_{u,V}  + \frac{211}{399} C_{d,V }\right|  $ &
$\leq 1.8 \times 10^{-12}  $ &
\cite{Porsev:2009pr,Arcadi:2019uif}  \\
 $ |2C_{u,A}+C_{d,A}| \left| \frac{188}{399}  C_{u,V}  + \frac{211}{399} C_{d,V }\right|  $ &
$\leq 2.0 \times 10^{-8}  $ &
\cite{Dzuba:2017puc}  \\
 $ |2C_{u,V}+ C_{d,V}| $ &
$\leq 3.6 \times 10^{-4} / \sqrt{{\rm BR}(Z' \to e e)}$ &
\cite{NA482:2015wmo} \\
 $ \sqrt{C^2_{e,V}+ C^2_{e,A}} $ &
$\leq 1 \times 10^{-5} / \sqrt{{\rm BR}(Z' \to \nu \nu)}$ &
\cite{NA64:2023wbi} \\

\hline
\end{tabular}
\caption{Summary of the experimental constraints used.}
\label{tab:constraints}
\end{table}

\section{The experimental framework sensitive to the \xst\ }
\label{Sect3}

In order to analyse the \cevns\ data from reactors we need to calculate the nuclear recoil spectrum in the SM as well as when including an extra vector boson exchange such as the \xst.
Following the same general framework as in~\cite{Cederkall:2025bka}, with $y=E_r/E_r^{\max}$ being the fractional nuclear recoil energy and $x=E_\nu/E_\nu^{\max}$ being the fractional neutrino energy, we write the theoretical recoil spectrum as
\begin{eqnarray}
\label{eq_dndy}
 \dfrac{d N_r}{dy} & = & \dfrac{ m_{\rm det} t_{\rm exp}}{M_{\rm A}} N_{\rm A}   \int_{\sqrt{y}}^{1}  \dfrac{d \sigma^{\nu_eN}}{dy}  \dfrac{d \Phi_{\overline{\nu}_e}}{dx} dx, 
\end{eqnarray}
where $m_{\rm det}$ is the mass of the detector, $t_{\rm exp}$ is the exposure time, $M_{\rm A}$ is the mass of the detector nucleus (in kg/kmol), $ N_{\rm A}$ is the Avogadro's number, $\frac{d\sigma^{\nu_eN}}{dy}$ is the differential cross-section for coherent elastic $\nu_eN$-scattering
and
$ \frac{d \Phi_{\overline{\nu}_e}}{dx}$ is the neutrino flux from a reactor. In order to have the same formalism for neutrinos from reactors and $\pi^+$ decays at rest, we use $E_\nu^{\max}=m_\mu/2$ and $E_r^{\max}=m_\mu^2/(2M)$. For a Germanium detector, as used both by the CONUS+ and Dresden-II experiments,  with $M=67.6$ GeV it then follows that $E_r^{\max}=82.5$ keV.

Using the same formalism as in~\cite{Cederkall:2025bka}, the \cevns\ cross-section including an additional $Z^\prime$ boson with mass $m_{Z^\prime}$ for electron or muon neutrinos,  can be written as
\begin{eqnarray}
\label{eq_dsigdy}
\dfrac{d\sigma^{\nu_{e/\mu}N}}{dy} & = & \dfrac{ [ N - (1-4\sin^2\theta_W)Z]^2 \, [F_V(y)]^2}{4\pi m_\mu^2}  \, \left(1- \dfrac{y}{x^2} \right)
\left(  \dfrac{G_Fm_\mu^2}{\sqrt{2} }  - \dfrac{C^{\nu_{e/\mu}}_{\rm eff}  }{ y + m_{Z^\prime}^2/m_\mu^2}   \right)^2,
\end{eqnarray}
with the effective $\nu_{e/\mu}N$ coupling $C^{\nu_{e/\mu}}_{\rm eff}$ mediated by the $Z^\prime$ given by the vector couplings of the $Z^\prime$ to the neutrinos ($C_{{\nu_{e/\mu}},V}$) and quarks ($C_{d/u,V}$)  as
\begin{eqnarray}
\label{eq_EffectiveNeutrinoCoupling}
C^{\nu_{e/\mu}}_{\rm eff} &=&\dfrac{C_{{\nu_{e/\mu}},V} [( N+2Z)C_{u,V} + (2N+Z) C_{d,V}] }{N - (1-4\sin^2\theta_W)Z} 
\end{eqnarray}
and $F_V(y)$ being the form factor for which we use the Klein-Nystrand one~\cite{Klein:1999qj},
\begin{equation}
F_{V}(y)  = \dfrac{3 \,  j_1(\sqrt{y}m_\mu R_A)}{\sqrt{y}m_\mu R_A}\dfrac{1}{1+a^2ym_\mu^2},
\end{equation}
where $j_1$ is a spherical Bessel function of the first kind, i.e., {$j_1(z)=(\sin z-z\cos z)/z^2$}, $R_A =1.2 A^{1/3} $ fm, $A=72.6$ and $a=0.7$ fm.

\begin{figure}[th]
\begin{center}
\includegraphics[width=8cm]{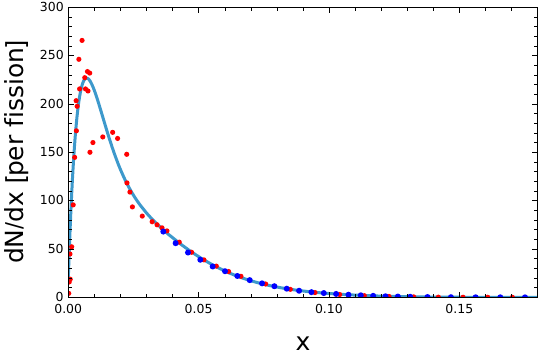}
\includegraphics[width=8cm]{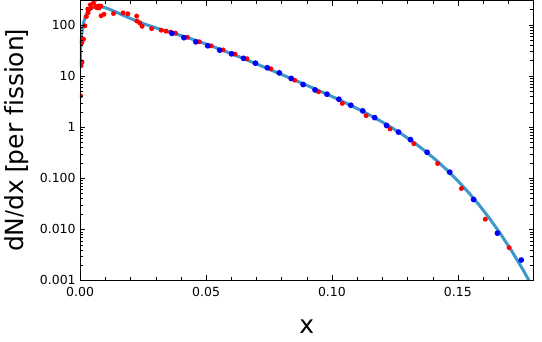}
\end{center}
\vspace*{-0.5cm}
\caption{Neutrino spectra from fission reactors as calculated by Kopeikin (in red dots) together with data from Daya Bay (in blue dots) and fit used (solid line). The spectra are normalised per fission.}
\label{fig_neutrinospectra}
\end{figure}

\begin{figure}[th]
\begin{center}
\includegraphics[width=8cm]{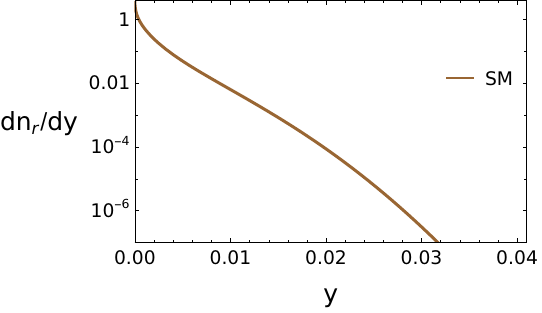} 
\end{center}
\vspace*{-0.5cm}
\caption{Nuclear recoil spectrum in the SM from reactor neutrinos and assuming a Germanium detector before smearing and quenching. Arbitrary normalisation.}
\label{fig_recoilspectra}
\end{figure}

The reactor neutrino spectra have been calculated by Kopeikin~\cite{Kopeikin:2012zz} as well as by Mueller {\it et al.}~\cite{Mueller:2011nm} and Huber~\cite{Huber:2011wv}. 
In the following we also use the Daya Bay measurement at high energies~\cite{DayaBay:2022eyy,DayaBay:2021dqj}.
As a first step towards modelling the neutrino flux we parametrise the latter per fission from a reactor as 
\begin{equation}
 \dfrac{d N_{\overline{\nu}_e}}{dx} = e^{a_1 + b_1 x + c_1 x^2 + d_1 x^3 + e_1 x^4 + f_1 x^5}+ xe^{a_2 + b_2 x + c_2 x^2},
\end{equation}
where $a_1 = 2.87593$, $b_1 = 186.145$, $c_1 = -3358.36$, $d_1 = 31940.8$, $e_1 = -139054$, $f_1 = 193450$, $a_2 = 11.2864$, $b_2 = -149.915$ and $c_2 = -452.33$.
The parameters have been obtained by fitting to the Daya Bay data for $x>0.0379$ ($E_\nu > 2$ MeV) and then subsequently to the calculation by Kopeikin. The resulting fit is also shown in Fig.~\ref{fig_neutrinospectra} together with the data used. The neutrino flux  from a given reactor is then obtained as 
\begin{equation}
 \dfrac{d \Phi_{\overline{\nu}_e}}{dx} =   \dfrac{ \Phi}{N_{\overline{\nu}_e}}  \dfrac{d N_{\overline{\nu}_e}}{dx},
\end{equation}
where the total fluxes for the Dresden-II and CONUS+ experiments are:
\begin{itemize}
\item
$\Phi_{\rm Dresden-II} = 4.8 \times 10^{13}$ cm$^2/$s~\cite{Colaresi:2022obx},
\item
$\Phi_{\rm CONUS+} = 1.5 \times 10^{13}$ cm$^2/$s~\cite{Ackermann:2025obx},
\end{itemize}
with the number of neutrinos per fission $N_{\overline{\nu}_e}= 6.8$,  from the parametrisation in agreement with~\cite{Kopeikin:2012zz}. For reference, the resulting recoil spectrum in the SM is shown in Fig.~\ref{fig_recoilspectra}. From the figure it is clear that the spectrum resulting from reactor neutrinos is very steeply falling compared to the one from $\pi^+$ decays at rest.

Note that a given nuclear recoil energy requires a minimum neutrino energy which is reflected by the constraint $x>\sqrt{y}$ in Eq.~(\ref{eq_dndy}).  For example, neutrinos with at least  2 MeV ($x=0.0379$) are needed to contribute to the recoil spectrum at $E_r$ = 0.118 keV ($y=0.0014$) for a Germanium detector. Therefore, the modelling of the shape of the neutrino energy spectrum at small $x$ is not so important, as long as the total number of neutrinos per fission is correct, since the shape of the spectrum in this region is determined by the energy resolution of the detector (as it will become clear below).

From the theoretical recoil spectrum in Eq.~(\ref{eq_dndy}), the experimentally observed one is obtained by taking into account the Quenching Factor (QF), which describes the conversion of the nuclear recoil energy to an ionisation signal, as well as the finite detector resolution giving
\begin{eqnarray}
\label{eq_dndyrec}
\dfrac{dN_r}{dy_{\rm rec}} & = &  \int_{y_{\rm min}}^1 R\left(y_{\rm rec},y_{\rm ion}\right) \dfrac{dN_r}{dy}  dy \, ,
\end{eqnarray}
where $y_{\rm rec}$ is the reconstructed energy fraction,  $y_{\rm ion}$ is the fractional ionisation energy ($y_{\rm ion}=E_{\rm ion}/E_r^{\max}$) and $ R\left(y_{\rm rec},y_{\rm ion}\right)$ describes the detector response in terms of  the resolution, which we define below. The lower integration limit, $y_{\rm min}= \dfrac{\eta}{E_r^{\max}} \dfrac{1}{Q(y_{\rm min})}$, is the minimal fractional ionisation energy in terms of the minimal ionisation energy $\eta=2.96$ eV$_{\rm ee}$ needed to create an electron-hole pair in Germanium and $Q(y)$ is the QF.

\begin{figure}[th]
\begin{center}
\includegraphics[width=8cm]{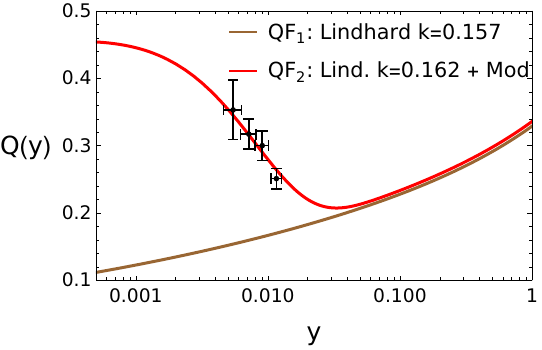} 
\end{center}
\vspace*{-0.5cm}
\caption{QFs used: ${\rm QF}_1$ is the Lindhard model with $k=0.157$ and ${\rm QF}_2$ is the Lindhard model with $k=0.162$ and a modification added to describe the data points.}
\label{fig_qfactor}
\end{figure}

In the following we mainly use the Lindhard model to describe the quenching. In this model the fractional ionisation energy created by the nuclear recoil is written as
\begin{equation}
    y_{\rm ion}  = Q(y)y,
\end{equation}
where
\begin{equation}
    Q_{\rm Lindhard}(y) = \dfrac{kg(y)}{1 + kg(y)}
\end{equation}
and
$g(y) = 3(C_Z y)^{0.15} + 0.7(C_Z y)^{0.60}  + C_Z y $ with $C_Z = E_r^{\rm max} \times 11.5/Z^{7/3}$ and $k = 0.157$. The resulting QF is shown in Fig.~\ref{fig_qfactor} together with a modified QF which has been constructed to describe data from \cite{Collar:2021fcl} which are also shown in the figure. 

To describe these data we simply add the modification 
\begin{equation}
    Q_{\rm Mod}(y) = 0.36 \exp(-120 y)
\end{equation}
to the QF. In addition, in order to illustrate the uncertainties related to the QF we also modify the factor $k$ to $0.162$. Thus we consider the two following QFs
\begin{eqnarray}
{\rm QF}_1& = & Q_{\rm Lindhard}(k=0.157), \\
{\rm QF}_2& = & Q_{\rm Lindhard}(k=0.162) + Q_{\rm Mod}.
\end{eqnarray}
With the two QFs defined, we then also get the corresponding minimal fractional recoil energies in Eq.~(\ref{eq_dndyrec}).
For the Lindhard model ${\rm QF}_1$ this becomes  $y_{\rm min}= 0.00034$ whereas for the modified Lindhard model  ${\rm QF}_2$ it is given by $y_{\rm min}= 0.000083$. 

Next we consider the effects of the finite detector resolution as described by
\begin{equation}
\label{eq_ge_resolution}
R(y_{\rm rec},y_{\rm ion})=\dfrac{2}{1+ {\rm Erf }(\frac{y_{\rm ion} }{\sqrt{2} \sigma})} \dfrac{1}{\sqrt{2\pi} \sigma } \exp\left[ - \dfrac{(y_{\rm rec} - y_{\rm ion})^2}{2\sigma^2 }\right].
\end{equation}
where $\sigma=\sigma_E/E_r^{\max}$ is a dimensionless width with
\begin{equation}
\sigma_E^2 =  \sigma_n^2 + E_{\rm ion} \eta F,  
\end{equation}
where $\eta=2.96$ eV$_{\rm ee}$, the Fano factor $F=0.11$ and $\sigma_n$ is the intrinsic resolution of the detector for which we use
\begin{eqnarray}
    \sigma_n^{\rm Dresden-II} &=& 68.5 \, {\rm  eV}_{\rm ee},\\
    \sigma_n^{\rm CONUS+} &=& 20.0  \, {\rm  eV}_{\rm ee}.
\end{eqnarray}

Note that the transformation above from $\dfrac{dN_r}{dy} $ to $\dfrac{dN_r}{dy_{\rm rec}} $ preserves the total number of events if the resolution $\sigma$ is changed, as long as  $y_{\rm min}$ is kept constant and the factor $2/({1+ {\rm Erf }(\frac{y_{\rm ion} }{\sqrt{2} \sigma})})$, which compensates for the Gaussian distribution being cut off, is included.  In other words:
\begin{equation}
\int_0^1 \dfrac{dN_r}{dy_{\rm rec}} dy_{\rm rec}  =  \int_0^1 \int_{y_{\rm min}}^1 R\left(y_{\rm rec},Q(y)y\right) \dfrac{dN_r}{dy}  \, dy \, dy_{\rm rec} =   \int_{y_{\rm min}}^1 \dfrac{dN_r}{dy}  dy. 
\end{equation}
However, since  $y_{\rm min}$ depends on the QF, the total number of events will change if the QF is changed in such a way that $y_{\rm min}$ changes.

\begin{figure}[th]
\begin{center}
\includegraphics[width=8cm]{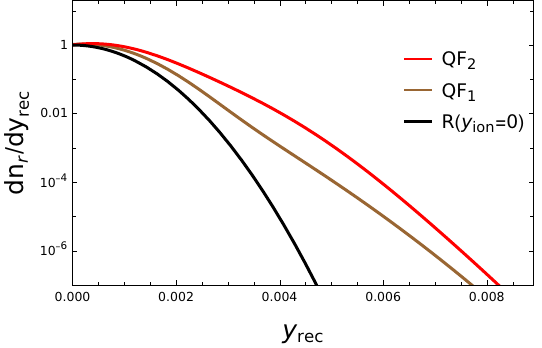} 
\includegraphics[width=8cm]{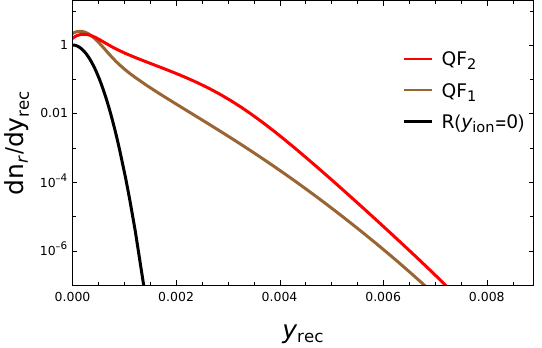} 
\end{center}
\vspace*{-0.5cm}
\caption{Shape of nuclear recoil spectra for  Dresden-II (left) and CONUS+ (right) after smearing and quenching with different QFs as well as the inherent resolution of the detector given by $R\left(y_{\rm rec},0\right)$ normalised such that $R\left(0,0\right)=1$.}
\label{fig_recoilspectrasmearedQF}
\end{figure}

The resulting detector signals for the Dresden-II and CONUS+ experiments are shown in Fig.~\ref{fig_recoilspectrasmearedQF}. Comparing to Fig.~\ref{fig_recoilspectra}, there are two main effects on the spectrum: (\textit{i}) the quenching pushes the spectrum to much smaller fractional recoil energies and (\textit{ii}) the finite detector resolution gives a substantial smearing of the spectrum - especially for the Dresden-II experiment - and in addition has a large impact on the normalisation in the region $y_{\rm rec} \gtrsim 0.002$ corresponding to $E_{\rm ion} \gtrsim 165$ eV$_{\rm ee}$. It is also clear that using different QFs has a very large impact on the spectral shape as well as normalisation. We finally note that a larger smearing gives higher sensitivity to smaller recoil energies.

\begin{figure}[th]
\begin{center}
\includegraphics[width=8cm]{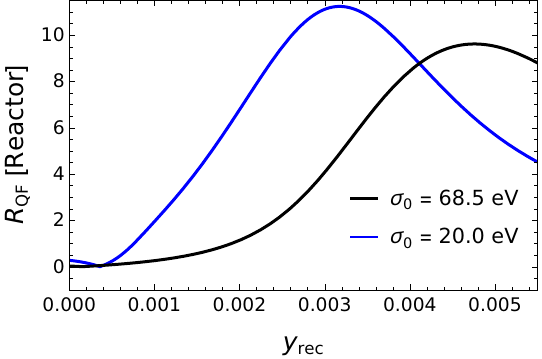} 
\includegraphics[width=8cm]{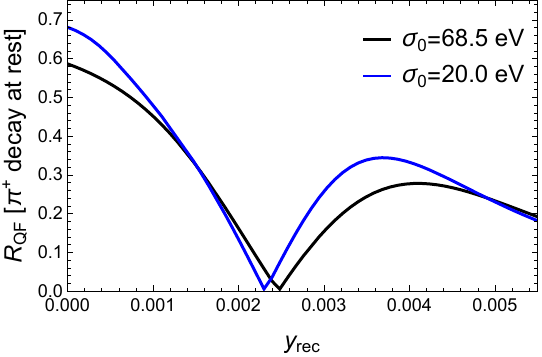}
\end{center}
\vspace*{-0.5cm}
\caption{QF uncertainty for reactor (left) and $\pi^+$ decay at rest (right) neutrinos with the same detector resolution as Dresden-II (black) and CONUS+ (blue).}
\label{fig_quenchinguncertainty}
\end{figure}

In order to quantify the uncertainty from quenching, we define 
the QF uncertainty  by calculating the ratio
 \begin{equation}
 \label{eq_QFuncertainty}
 R_{\rm QF} = \left|\frac{\left.\dfrac{dN_r}{dy_{\rm rec}}\right|_{\rm QF2} -\left.\dfrac{dN_r}{dy_{\rm rec}}\right|_{\rm QF1}}{ \left.\dfrac{dN_r}{dy_{\rm rec}}\right|_{\rm QF1} } \right|.
 \end{equation}
As can be seen in Fig.~\ref{fig_quenchinguncertainty}, the uncertainty defined in this way reaches about a factor 10 as its maximal value in the region where there are experimental data from reactors. In addition, the peak uncertainty is obtained for quite different values of the detector signal, as it depends on the intrinsic resolution of the detector. For comparison, the figure also shows the same uncertainty calculated for  $\pi^+$ decay at rest, showing that in this case the uncertainty is more than an order of magnitude smaller above the detector threshold of about $y_{\rm rec} \approx 0.002$ ($E_{\rm rec} \approx 160$ eV$_{\rm ee}$).

To convert the calculations done so far to the number of events counted as function of energy, which can be compared to experiment, we need the detector masses and the exposure time. For CONUS+ the detector mass is 2.83 kg and the corresponding exposure of 115.5 days giving a total of 327 kg days whereas for Dresden-II the mass is 2.924 kg and the exposure 96.4 days giving 282 kg days.

With the above we get the predicted number of nuclear recoil events for CONUS+ in the SM (integrating from $y_{\rm rec}=0.160/82.5$ to $y_{\rm rec}=0.400/82.5$) to be $381$ with ${\rm QF}_1$ and 3622 with ${\rm QF}_2$, which should be compared to the collaboration's calculation of $347\pm59$ and the observation of $395\pm106$, whereas for the Dresden-II experiment we get, integrating from $y_{\rm rec}=0.200/82.5$ to $y_{\rm rec}=0.400/82.5$, 2245 nuclear recoil events with  ${\rm QF}_1$ and 9817 with  ${\rm QF}_2$. Note that even though the detector masses and exposure times are quite similar, the CONUS+ sees much fewer events due to the flux being a factor three smaller and the resolution being higher.

\begin{figure}[th]
\begin{center}
\includegraphics[width=8cm]{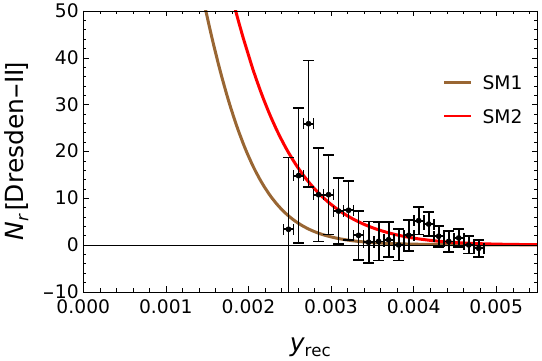} 
\includegraphics[width=8cm]{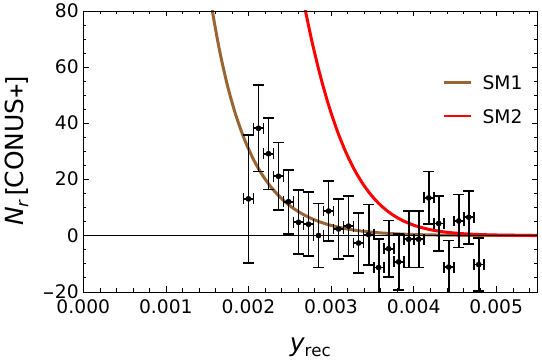} 
\end{center}
\vspace*{-0.5cm}
\caption{Nuclear recoil spectra after smearing and quenching compared to data as published using the QF1 or QF2 in the SM. The data has been normalised in the same way as done by the respective experiments.}
\label{fig_recoilspectrasmeareddata}
\end{figure}

Comparing to the measured recoil spectra, as shown in Fig.~\ref{fig_recoilspectrasmeareddata}, we see that the CONUS+ data are in very good agreement with the SM for ${\rm QF}_1$ but not at all with ${\rm QF}_2$, whereas the Dresden-II data seems to be in much better agreement with the SM when using ${\rm QF}_2$. 
It has recently been pointed out~\cite{Li:2025pfw} that no QF can be found making the two datasets agree with each other. In the following we show that on the one hand the datasets can be made to agree with each other in the SM by including the QF uncertainty as defined in Eq.~(\ref{eq_QFuncertainty}) and, more interestingly, we also show that the datasets can be made to agree with each other using the standard Lindhard model and adding the \xst\ with appropriate couplings.

\subsection{Statistical and systematic errors}

In order to quantify these statements in more detail we use the following $\chi^2$ function to analyse the statistical significance of the data:
\begin{equation}
\label{eq_chi2}
\chi^2=\sum_i \left(\dfrac{(1+\rho)x_i-\mu_i}{\sigma_i}\right)^2+ \left(\dfrac{\rho}{\sigma_{\rm sys}}\right)^2 \, ,
\end{equation}
where $x_i$ are the model predictions for the number of events in a bin and $\mu_i$ are the number of observed events with $\sigma_i$ being the corresponding errors. In addition, $\rho$ is an overall scale factor, with an overall systematic uncertainty $\sigma_{\rm sys}$. The scale factor is determined by minimising the $\chi^2$ with respect to $\rho$ giving
\begin{equation}
\rho= \dfrac{\sum_i \dfrac{x_i(\mu_i-x_i)}{\sigma_i^2}}{ \dfrac{1}{\sigma_{\rm sys}^2}+\sum_i \dfrac{x_i^2}{\sigma_i^2} } \, .
\end{equation}
Any systematic uncertainties which vary from bin to bin, $\sigma_{{\rm sys}, i}$, are included  in the errors $\sigma_i$ using  $\sigma_i^2 = \sigma_{{\rm stat},i}^2 + \sigma_{{\rm sys},i}^2x_i^2$, which corresponds to including a scale factor $\rho_i$ for each bin where, in addition, we use the SM values for $x_i$ for simplicity. To be more clear, when including the QF uncertainty from Eq.~(\ref{eq_QFuncertainty}), we calculate the average $R_{\rm QF,i}$ for each bin and then add them quadratically to the errors reported by the experiments setting $\sigma_{{\rm sys},i}=R_{\rm QF,i}$.

When analysing the reactor datasets, we will for definiteness use the same  scale uncertainties for both of them: $\sigma_{\rm sys}=0.15$ when the QF uncertainty is handled separately for each bin and  $\sigma_{\rm sys}=0.17$ when it is included as part of the overall uncertainty following~\cite{Ackermann:2025obx}.

With $\chi^2$ defined as in Eq.~(\ref{eq_chi2}) we get $\chi^2=11.6 \,(37.4)$ for CONUS+ with ${\rm QF}_1$(${\rm QF}_2$) and $\chi^2=14.4 \,  (8.2)$ for Dresden-II with ${\rm QF}_1$(${\rm QF}_2$) in the SM when not taking into account the QF uncertainty bin-by-bin. Adding the latter, the $\chi^2$ values change to $\chi^2= 7.2$ for CONUS+  and $\chi^2=10.9$ for Dresden-II, in both cases with ${\rm QF}_1$.   Comparing these $\chi^2$ values with the number of data points used in the calculation, 24 for CONUS+ and 20 for Dresden-II, we see that the $\chi^2$ per degree of freedom is well below one except when using ${\rm QF}_2$ for the CONUS+ data and not including the QF uncertainty. Based on these observations, we use ${\rm QF}_1$, with or without the QF uncertainty, as the main model for analysing the data.

\begin{figure}[th]
\begin{center}
\includegraphics[width=7.cm]{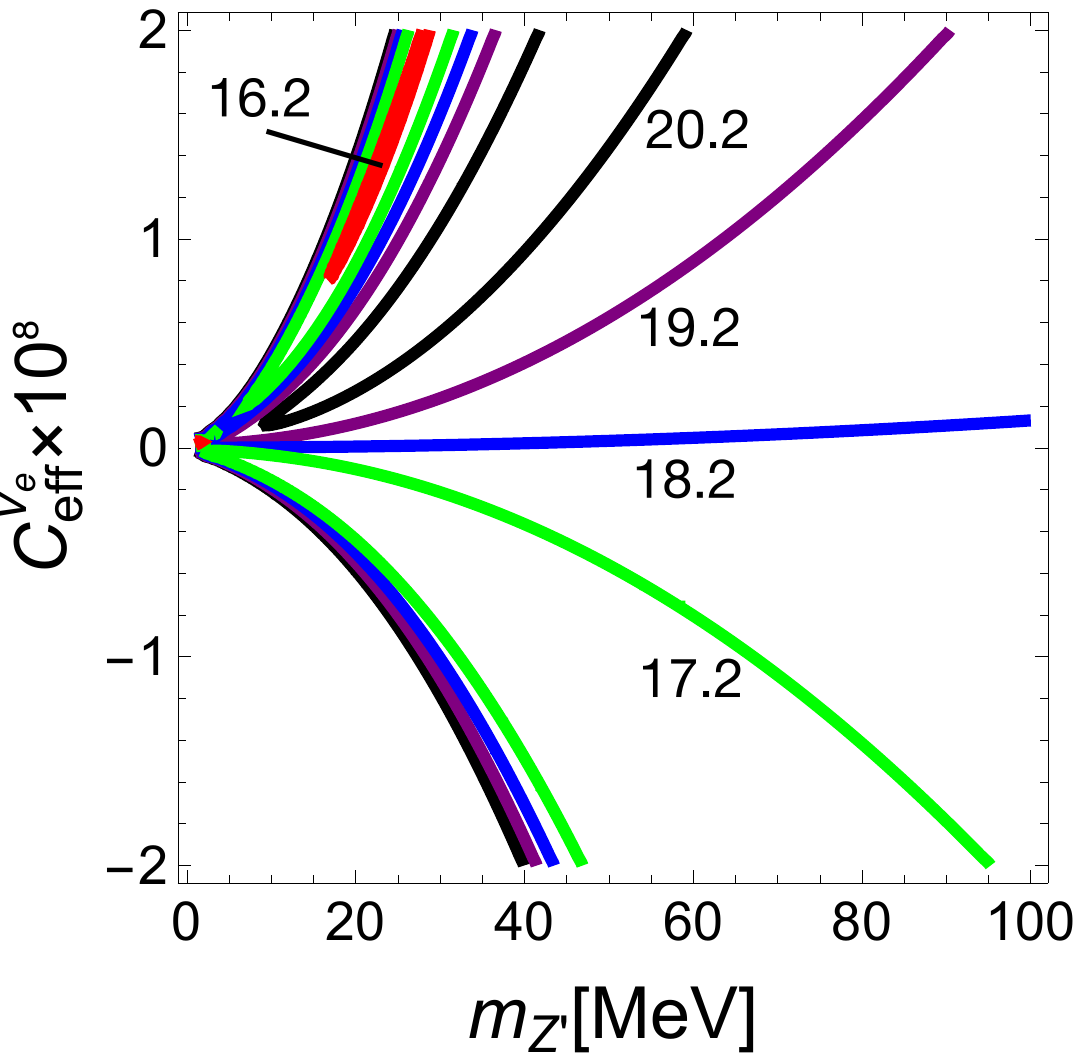} 
\hspace*{0.5cm}
\includegraphics[width=7.cm]{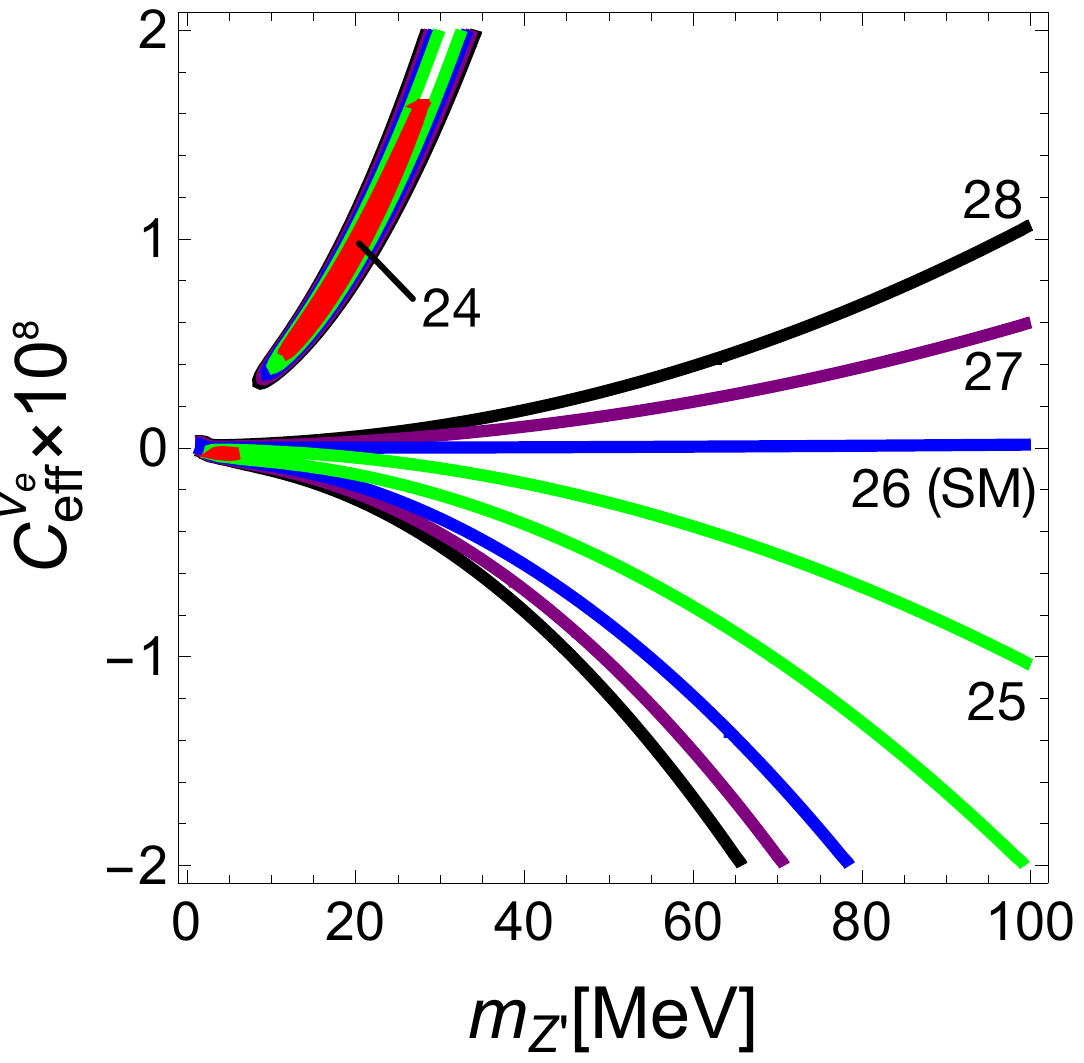} 
\end{center}
\vspace*{-0.5cm}
\caption{The $\chi^2$ contours for  combined reactor datasets  with (left) and without (right)  bin-by-bin QF uncertainty added to data.}
\label{fig_contour_reactor}
\end{figure}

\begin{figure}[th]
\begin{center}
\includegraphics[width=8cm]{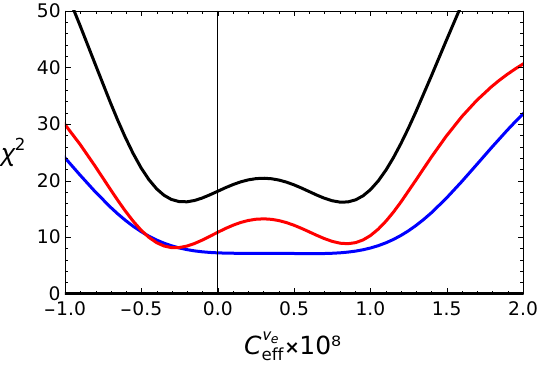}
\includegraphics[width=8cm]{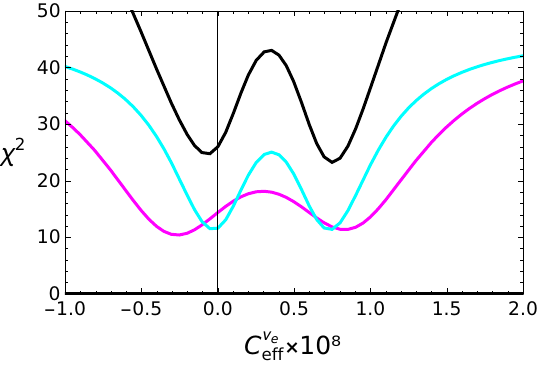} 
\end{center}
\vspace*{-0.5cm}
\caption{The $\chi^2$ for the individual (magenta, red = Dresden-II and cyan, blue = CONUS+) as well as combined (black) reactor datasets: with (left) and without (right) QF uncertainty added to data for $m_{Z^\prime}=16.7$ MeV.}
\label{fig_chi2}
\end{figure}

Interestingly, it is also possible to explain the apparent differences between the CONUS+ and Dresden-II experiments by including an additional light $Z^\prime$ as defined by the cross-section in Eq.~(\ref{eq_dsigdy}). A priori, such an additional light $Z^\prime$ could have any mass and effective coupling to nuclei. We have therefore calculated the $\chi^2$ when varying these two parameters to see what ranges are preferred/allowed by the reactor data. The results are shown in Fig.~\ref{fig_contour_reactor}. From the figure, it is clear that with the QF uncertainty added there is a range of allowed $C^{\nu_{e}}_{\rm eff}$ values for any given $m_{Z^\prime}$. For larger masses the limits scale as $C^{\nu_{e}}_{\rm eff}/m_{Z^\prime}^2$ as is evident from  Eq.~(\ref{eq_dsigdy}). At the same time, when the QF uncertainty is not added, there is a preferred region $m_{Z^\prime}\approx [10,30]$ MeV and  $C^{\nu_{e}}_{\rm eff}\approx[0.5,1.5]\cdot 10^{-8}$ which has $\chi^2 \approx 24$ compared to the SM value 26. More specifically, for  $m_{Z^\prime} = 16.7$ MeV, the $\chi^2$ for different effective neutrino nucleus couplings is shown in Fig.~\ref{fig_chi2}. Here it is clear that the data are better described by including a 16.7 MeV $Z^\prime$ with  $C^{\nu_{e}}_{\rm eff}$ being either slightly negative or close to $[0.7,0.9]\cdot 10^{-8}$ irrespectively of whether the QF uncertainty is taken into account or not.

\subsection{Combining reactor and spallation source datasets}

Before comparing the reactor datasets to a specific \xst\ model, we want to include additional data on \cevns\ from the COHERENT experiment. We use both the Argon~\cite{COHERENT:2020iec} and Caesium Iodine~\cite{COHERENT:2021xmm} data as well as the recent Germanium data~\cite{COHERENT:2024axu}. We use the same framework for analysing these data as for the reactor cases, although generalised to include also muon neutrinos. In short, the cross-section for $\nu_{e/\mu}$ nucleus scattering is given by Eq.~(\ref{eq_dsigdy}) and the neutrino fluxes from the $\pi^+$ decay at rest are
\begin{eqnarray}
 \dfrac{d\Phi_{\nu_e}}{dx} & = & \dfrac{r N_{\mathrm{POT}}}{4\pi L^2}  \,  12x^2(1-x) \label{eq_nue-flux},\\
  \dfrac{d\Phi_{\bar{\nu}_\mu}}{dx} & = & \dfrac{r N_{\mathrm{POT}}}{4\pi L^2}  \,  2x^2(3-2x)   \label{eq_numubar-flux},\\
  \dfrac{d\Phi_{\nu_\mu}}{dx} & = &  \dfrac{r N_{\mathrm{POT}}}{4\pi L^2} \, \delta \left( x- x_0 \right), \label{eq_numu-flux}
\end{eqnarray}
where $r$ is the number of $\pi^+$ produced per proton, $N_{\mathrm{POT}}$ is the number of protons on target, $L$ is the distance travelled by the neutrinos and 
\begin{equation}
x_0 = \dfrac{m_{\pi^+}^2-m_\mu^2}{m_\mu m_{\pi^+}} \approx 0.564 \, 
\end{equation}
is the scaled energy of the muon neutrinos from the two-body decay.
Convoluting the cross-section with the neutrino fluxes, the theoretical nuclear recoil spectrum for neutrinos from pion decays at rest is then given by
\begin{eqnarray}
\label{eq_recoilspec}
\dfrac{dN_r}{dy} & = & \dfrac{r N_{\rm POT}}{4 \pi L^2} \dfrac{1}{2\pi m_\mu^2} [ N - (1-4\sin^2\theta_W)Z]^2  \dfrac{ m_{\rm det}}{M_{\rm A}} N_{\rm A} \, [F_V(y)]^2 \\ \nonumber 
&& \left\{ \left(  \dfrac{G_Fm_\mu^2}{\sqrt{2} }  - \dfrac{C^{\nu_e}_{\rm eff}  }{ y + m_{Z^\prime}^2/m_\mu^2}   \right)^2 \dfrac{dn_{\nu_e}}{dy}
+ \left( \dfrac{G_Fm_\mu^2}{\sqrt{2} } - \dfrac{C^{\nu_\mu}_{\rm eff}  }{ y + m_{Z^\prime}^2/m_\mu^2}   \right)^2 \dfrac{dn_{\nu_\mu}}{dy} 
\right\},
\end{eqnarray}
where 
\begin{eqnarray}
 \dfrac{dn_{\nu_e}}{dy} & = &   \frac{1}{2} - 3 y +4 y^{3/2}  -  \frac{3}{2} y^2\, , \\
 \dfrac{dn_{\nu_\mu}}{dy} & = &    \frac{1}{2} - 2 y +2 y^{3/2}  - \frac{1}{2} y^2 
 + \left(\dfrac{ 1}{2}- \dfrac{ y}{2x_0^2} \right) \Theta\left(1-\dfrac{y}{x_0^2} \right) \, .
\end{eqnarray}
For more details we refer to~\cite{Cederkall:2025bka}.

To analyse the COHERENT data we use the experimental conditions as published by the collaboration and summarised in appendix~\ref{app_coherentdata}. Similarly as when analysing the reactor data, we use ${\rm QF}_1$ for the Germanium data. However, the QF uncertainty is only considered for the reactor datasets. Thus, the systematic scale uncertainties in the different COHERENT datasets are as follows: $\sigma_{\rm sys}^{\rm Ar}=0.13$ (Argon),
 $\sigma_{\rm sys}^{\rm CsI}=0.12$ (Caesium Iodine),
 $\sigma_{\rm sys}^{\rm Ge}=0.103$ (Germanium).

The effective neutrino nucleus couplings, given by Eq.~(\ref{eq_EffectiveNeutrinoCoupling}),
include a dependence on the number of neutrons and protons. The main effect is that the $C_{d,V}$ coupling carries a $\sim$ 10\% larger weight than $C_{u,V}$ from $N$ being larger than $Z$ for the nuclei of interest. In detail, this dependence varies from nucleus to nucleus. In order to be able to only use two effective neutrino nucleus couplings we approximate this dependence by using $C^{\nu_{e/\mu}}_{\rm eff}({\rm Ar})=1.020 \, C^{\nu_{e/\mu}}_{\rm eff}({\rm Ge})$ and $C^{\nu_{e/\mu}}_{\rm eff}({\rm CsI})=0.953  \, C^{\nu_{e/\mu}}_{\rm eff}({\rm Ge})$, which are obtained by taking the average of the nuclear dependence of the effective neutrino quark couplings   defined analogously to Eq.~(\ref{eq_EffectiveNeutrinoCoupling}). 

\begin{figure}[!t]
\begin{center}
\includegraphics[width=10.cm]{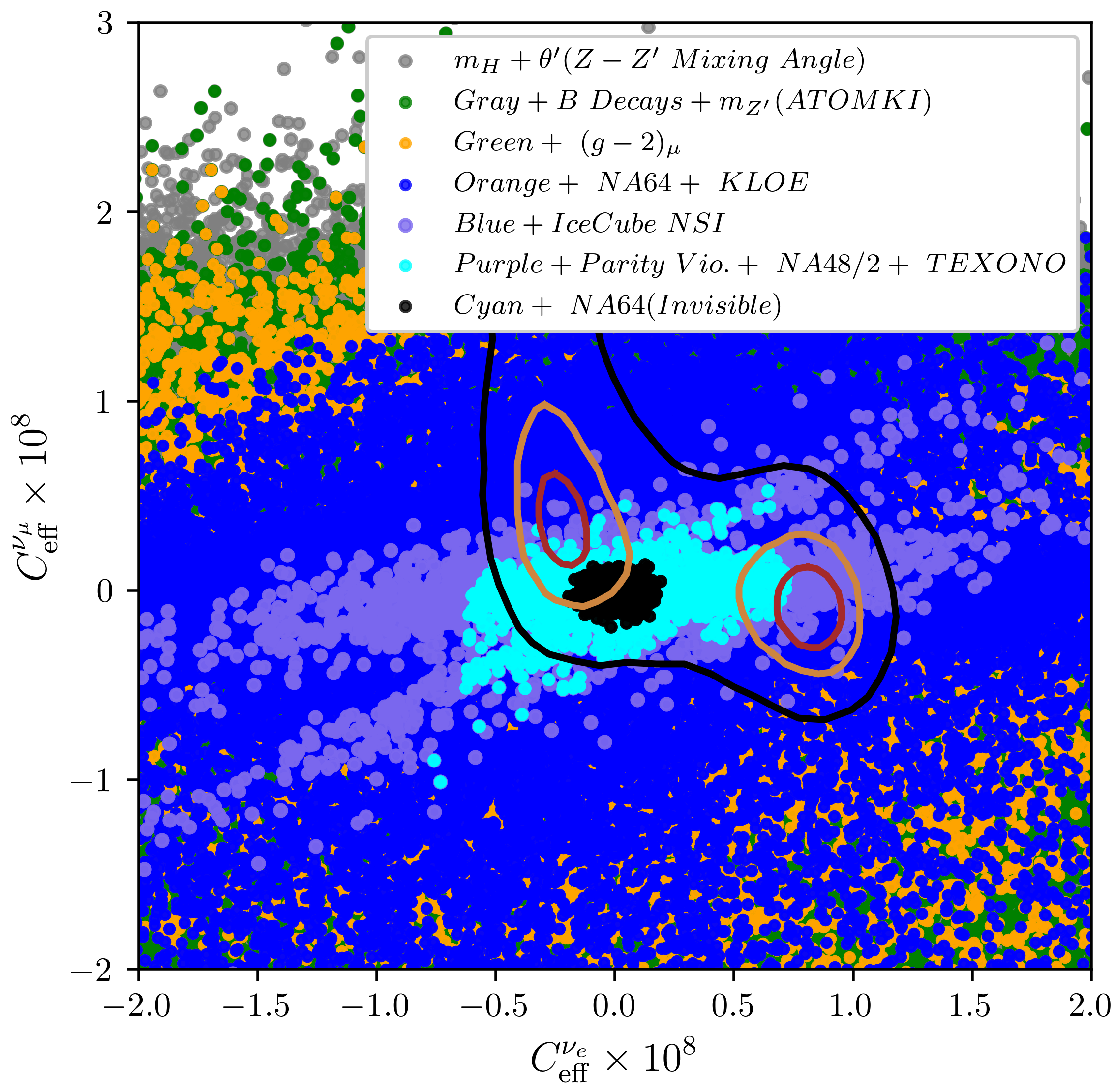} 
\includegraphics[width=10.cm]{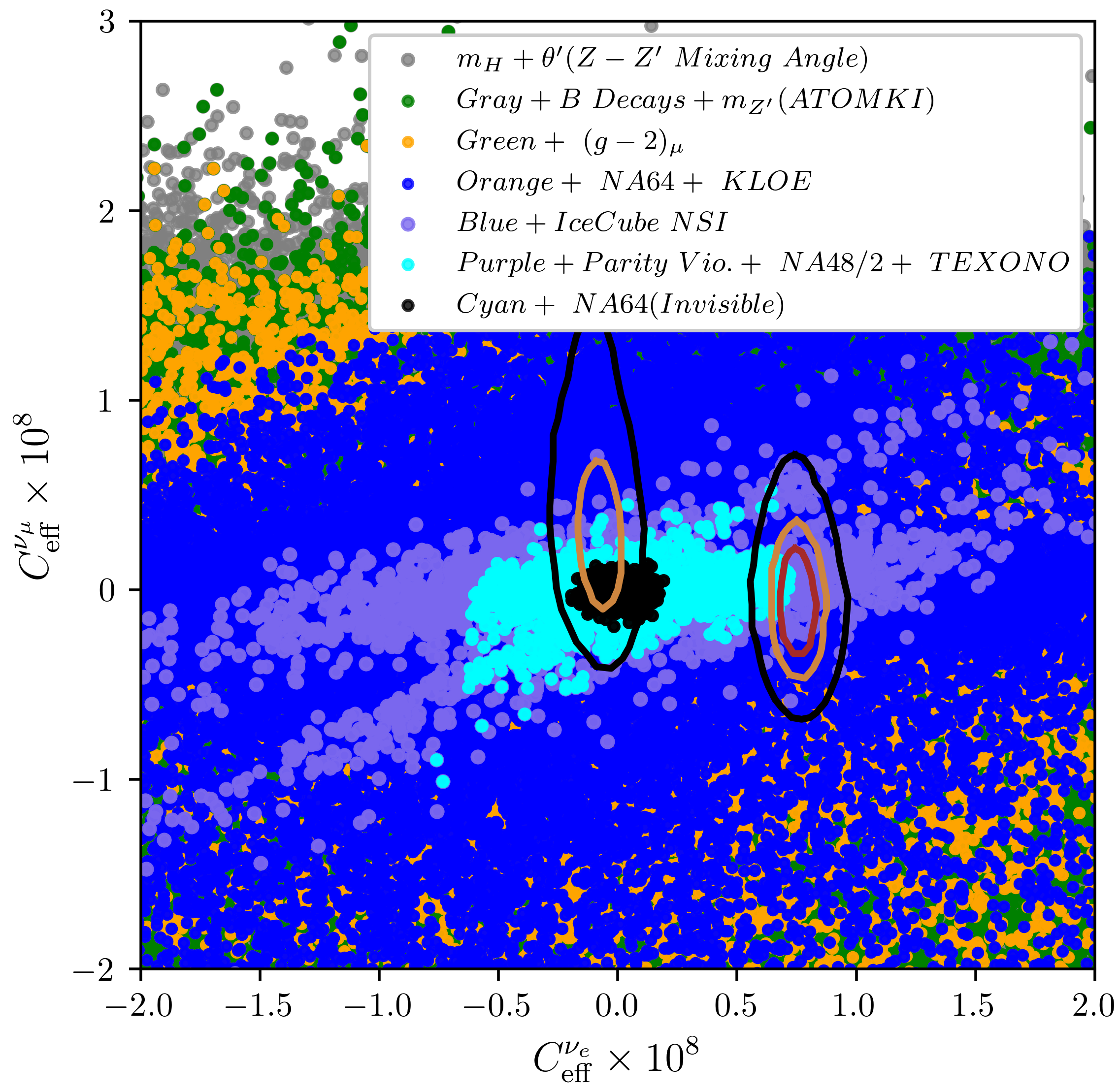} 
\end{center}
\vspace*{-0.5cm}
\caption{The $\chi^2$ contours for  combined reactor and COHERENT datasets  with (upper) and without (lower) QF uncertainty added to reactor data for $m_{Z^\prime}=16.7$ MeV. Brown denotes $\chi^2=\chi^2_{\rm SM}$, black $\chi^2=\chi^2_{\rm SM}+6$ and red  $\chi^2=\chi^2_{\rm SM}-2$. The coloured points show the results of the scan over parameters of the \xst\ model considered as defined in Tab.~\ref{tab_paramSP} after applying various constraints as indicated in the caption and detailed in Tab.~\ref{tab:constraints}.}
\label{fig_chi2AllStat}
\end{figure}

With these definitions, we can now calculate the experimentally observed recoil spectra also for the COHERENT data sets and compare to the published data using the $\chi^2$ function defined by Eq.~(\ref{eq_chi2}).
The resulting $\chi^2$ for $m_{Z^\prime}=16.7$ MeV and as a function of the effective neutrino nucleus couplings - $C^{\nu_{e}}_{\rm eff}$ and  $C^{\nu_{\mu}}_{\rm eff}$ - is shown in Fig.~\ref{fig_chi2AllStat}. As can be seen from the figure, adding  the COHERENT data does not change the preferred regions for $C^{\nu_{e}}_{\rm eff}$. At the same time, the COHERENT data constrains the $C^{\nu_{\mu}}_{\rm eff}$ coupling, although not quite as strongly as the reactor data constrain $C^{\nu_{e}}_{\rm eff}$.  It is also clear from the figure that the QF uncertainty does not have a big impact on the preferred regions even though the allowed regions are significantly larger when it is included.

The figure also shows the points resulting from a scan over the parameters of the \xst\ model presented in section~\ref{Sect2} and how they are constrained by the various experimental limits listed in Tab.~\ref{tab:constraints}. 
The most important ones are the NA64 and KLOE constraints, which limit the vector couplings of the electrons and, thereby, in this particular model, also constrain the vector couplings of the protons since the two are related, as is clear from Eq.~(\ref{eq:powers}). Next the NSI data from IceCube constrain the product of differences of vector couplings to neutrinos times the vector couplings to electrons, protons and neutrons, which further limits the allowed parameter space. The NA48/2 experiment constrains the vector couplings of protons even further and the TEXONO constraint limits the vector coupling to electron neutrinos given that the range of allowed vector coupling to the electron is very small. Finally, the NA64 constraint from DM searches gives very strong limitations on the couplings to neutrinos, essentially, $\sqrt{C_{\nu_e,V}^2+C_{\nu_e,A}^2+C_{\nu_\mu,V}^2+C_{\nu_\mu,A}^2+C_{\nu_\tau,V}^2+C_{\nu_\tau,A}^2} =\sqrt{2(C_{\nu_e,V}^2+C_{\nu_\mu,V}^2+C_{\nu_\tau,V}^2)} < 10^{-5}$, which, when combined with the NA48/2 constraint on the vector coupling of the proton, gives a strong limit in the model at hand wherein the vector coupling of the neutron is much smaller than that of the proton. Another possibility for interpreting the data, is "the protophobic" \xst-model, where instead the vector coupling of the neutron is much larger than that of the proton, which is analysed in some detail in our shorter paper~\cite{Rathsman:2026hif}. In such models the neutrino couplings are much smaller, such that the constraints from the DM searches at NA64 are fulfilled.

\begin{figure}[!t]
\begin{center}
\includegraphics[width=8.cm]{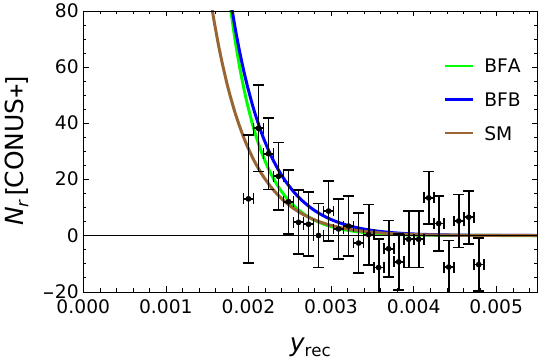} 
\includegraphics[width=8.cm]{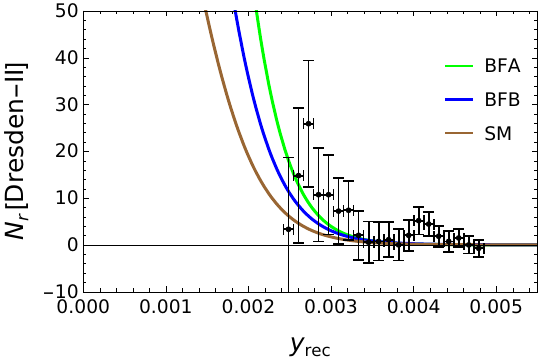} 
\includegraphics[width=8.cm]{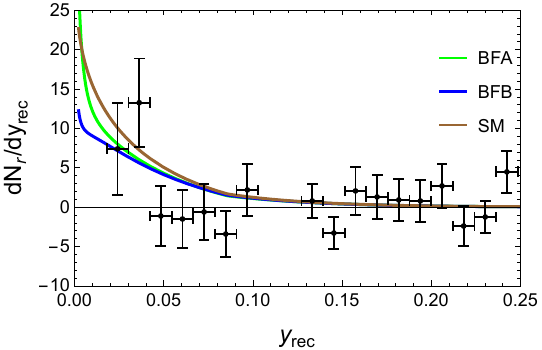} 
\includegraphics[width=8.cm]{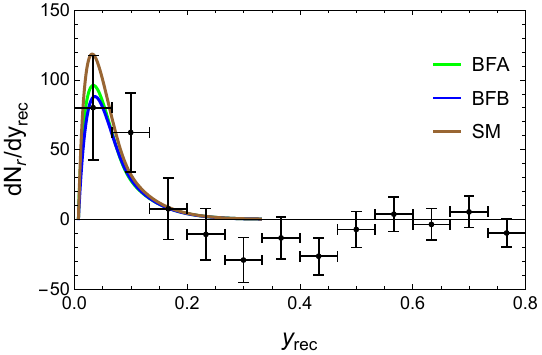} 
\includegraphics[width=8.cm]{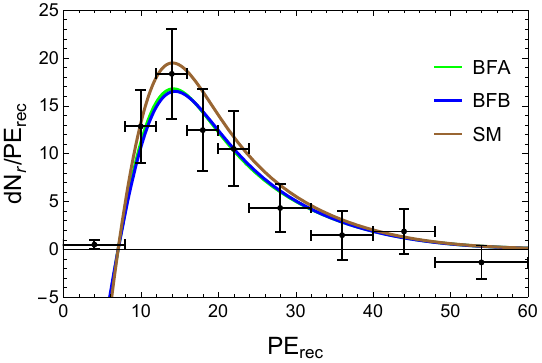} 
\end{center}
\vspace*{-0.5cm}
\caption{Nuclear recoil data from CONUS+ (upper left), Dresden-II (upper right), COHERENT Germanium (middle left), COHERENT Argon (middle right) and COHERENT CsI (lower) compared to the SM and best fit points A and B for $m_{Z^\prime}=16.7$ MeV.}
\label{fig_fitdata}
\end{figure}

All in all, from Fig.~\ref{fig_chi2AllStat}
we identify two distinct best fit points (or regions) {that are preferred by the \cevns\ data}, 
\begin{itemize}
    \item[] BFA: $C^{\nu_e}_{\rm eff} =0.8 \times 10^{-8}$ , $C^{\nu_\mu}_{\rm eff} = -0.1 \times 10^{-8}$,
    \item[] BFB:  $C^{\nu_e}_{\rm eff} =-0.2 \times 10^{-8}$ , $C^{\nu_\mu}_{\rm eff} = 0.3 \times 10^{-8}$, 
\end{itemize}
that we analyse in more detail below. 
It is also clear from from the results shown in the figure that applying all the constraints from other experiments only leaves the BFB region whereas the BFA region is not compatible with the NA64 constraints from DM searches. Even so, we analyse below how the two could be distinguished using only \cevns\ data. First of all, in 
Fig.~\ref{fig_fitdata}, we show the respective best fit points compared to the data used for the fit. As expected, both best fit points describe the data equally well. At the same time, it is clear that the spectra from the two fits differ at small nuclear recoils, below what has been measured by the COHERENT collaboration. Especially for the Argon and Caesium Iodine data, the detector efficiency cuts off the spectra for small recoil energies. For the reactor data, more data and a lowering of the detector threshold would be needed to be able to distinguish the two. 

\begin{figure}[th]
\begin{center}
\includegraphics[width=8cm]{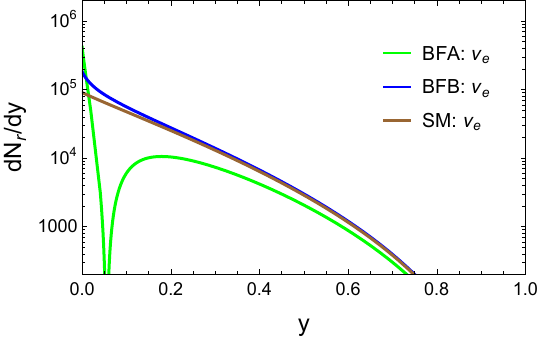}  
\includegraphics[width=8cm]{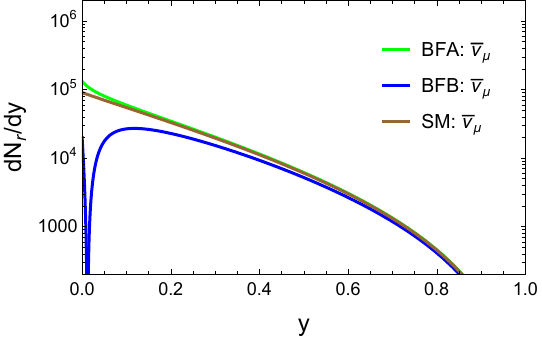} 
\includegraphics[width=8cm]{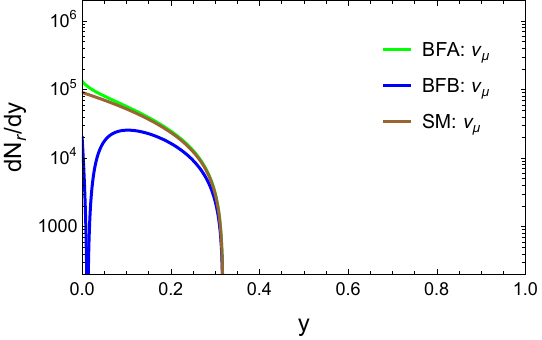} 
\includegraphics[width=8cm]{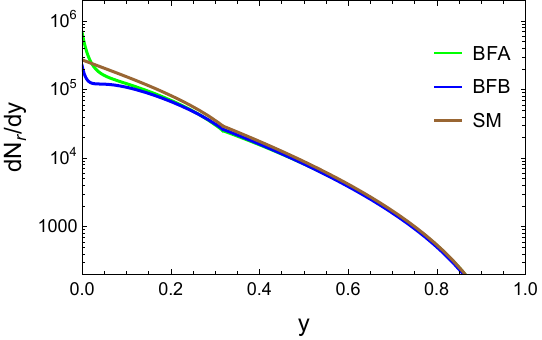} 
\end{center}
\vspace*{-0.5cm}
\caption{Theoretical recoil spectra for the best fit points A and B compared to SM for $m_{Z^\prime}=16.7$ MeV and assuming a Germanium detector. Upper left (right) shows the $\nu_e$ ($\overline{\nu}_\mu$) induced contribution whereas the lower left (right) plot shows the $\nu_\mu$ (sum of all) contribution(s).}
\label{fig_BMAB}
\end{figure}

To analyse in more detail the distinctions between the two best fit points, we show in Fig.~\ref{fig_BMAB} the theoretical recoil spectra for electron and (anti) muon neutrinos separately as well as the sum. From the figure it is clear that, in both BFA and BFB, the nuclear recoil spectrum for $\nu_e$-scattering is enhanced compared to the SM (upper left plot) at very small recoil energies, $y \lesssim  0.02$ whereas, at larger recoil energies, the spectrum is enhanced in a similar way in BFB and heavily suppressed in BFA. There is even a point where the recoil spectrum in BFA vanishes due to negative interference between the \xst\ and the SM.

\begin{figure}[th]
\begin{center}
\includegraphics[width=8cm]{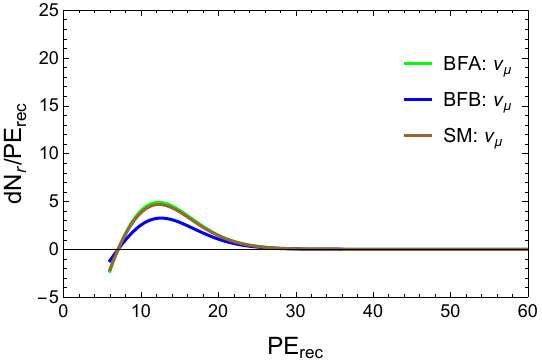} 
\includegraphics[width=8cm]{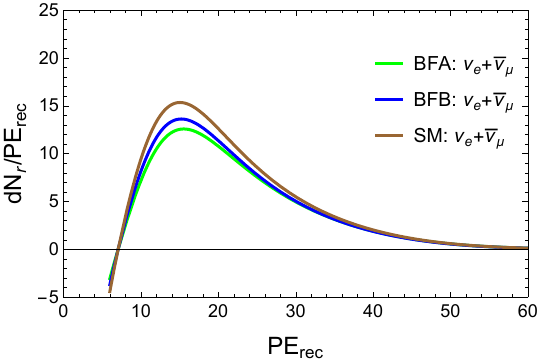} 
\end{center}
\vspace*{-0.5cm}
\caption{Recoil spectra after quenching and smearing for the best fit points A and B compared to SM for $m_{Z^\prime}=16.7$ MeV and assuming a CsI detector with the same characteristics as the COHERENT one. Left (right) prompt (non-prompt) contributions assuming that the timing information could be used to separate the two completely.}
\label{fig_BMAB_smeared}
\end{figure}

In order to fit the observations by COHERENT, this means that the  $\nu_\mu$ and $\overline{\nu}_\mu$-scattering (upper right and lower left plot respectively) induced recoil spectra at larger recoil energies, $y \gtrsim 0.2$, are enhanced in BFA and suppressed in BFB in such a way that the sum of all contributions (lower right plot) is the same. Thus, in order to separate the two scenarios, one needs to measure the recoil spectrum at lower energies than has been done at COHERENT: in other words, in the region $0.02 \lesssim y \lesssim 0.2$.

We then  also note that, in principle, the timing information from SNS events could help to distinguish the prompt  $\nu_\mu$ from the non-prompt  $\overline{\nu}_\mu$ and $\nu_e$ where the latter two have higher energies, but not to separate  $\nu_e$ from   $\overline{\nu}_\mu$. At the same time, as is clear from Fig.~\ref{fig_BMAB_smeared}, even with a perfect separation between  $\nu_\mu$ on the one hand and $\overline{\nu}_\mu$ and $\nu_e$ on the other, the limited acceptance at small recoil energies for the present COHERENT data makes the differences between the two scenarios very small as exemplified by the CsI results in the plot. So in the end it will be necessary to go to lower recoil energies to make full use of the timing information.

\section{Conclusions and outlook}
\label{Sect4}

By analysing nuclear recoil spectra arising from \cevns\ utilising electron anti-neutrinos from nuclear reactors and adding a light vector mediator in the form a $Z^\prime$ boson, we have shown that the data from the CONUS+ and Dresden-II experiments point to a preferred mass region (around 16.7 MeV) close to that of the \xst, which has emerged as a possible explanation of the ATOMKI anomaly. In fact,
including the \xst\ also makes it possible to resolve the tension between the CONUS+ and Dresden-II data. The difference in the observations between the two experiments is then to a large extent explained by the differences in detector resolution, which gives a different sensitivity to the \xst.

Assuming that, indeed, it is the \xst\ that explains this tension between the two experiments, we also find that the preferred effective coupling between the electron anti-neutrinos and nuclei mediated by the \xst\ is positive, in contrast to what is the case in the $B-L$ model and variations thereof. As such, this positive coupling is more in line with what is observed in the so-called universal models, which, however, do not fulfil the anomaly conditions.

In addition to the data from the two reactor experiments, which have observed \cevns, we also add data from the COHERENT experiment. When combined with the reactor data, these data also constrain the effective couplings between muon (anti-)neutrinos and nuclei mediated by the \xst. By making a combined fit to the data, we identified two regions of preferred effective neutrino-nucleus couplings for the \xst\ that are able to describe all available \cevns\ data from observations. 

From the analysis of the data it is also clear that models with non-universal flavour couplings are preferred. In fact, the preferred regions would not have been seen in a model with flavour universal couplings. Our analysis also shows that although currently the uncertainties from the quenching factor of Germanium is very large, the preferred regions of the effective neutrino-nucleus couplings mediated from the \xst\ are more or less the same irrespective of whether this uncertainty is included or not. In the end,
using a minimal anomaly-free model to add the \xst\ to the SM, we found that such a setup is indeed able to describe the \cevns\ data.  However, when adding  all currently known constraints from other  measurements, it turns out that only one of the regions preferred by the \cevns\ data is compatible with these constraints. In particular DM searches from NA64 put strong constraints on the particular model considered here.  

Going forward, our analysis also shows the importance of lowering the detector thresholds for experiments at spallation sources. We have also shown that adding timing information to distinguish prompt neutrinos from the $\pi^+$ decays and delayed ones from the $\mu^+$ decays could reveal dips in the recoil spectrum arising from negative interference between the $Z$ boson of the SM and the \xst. 

Finally, we note that there is a strong need to understand the quenching factors for Germanium detectors better in order to make full use of the \cevns\ data, in particular from reactors, for constraining BSM physics, such as the intriguing \xst.

\subsection*{Note added}
After finishing this paper, new results on the measurement of \cevns\  on a Germanium target by the COHERENT collaboration appeared~\cite{COHERENT:2026yje}. According to the collaboration, these data are more in line with the SM expectations. Such a change would have only a minor impact on our analysis, possibly resulting in a slight change of the preferred value of the effective muon-neutrino nucleus coupling.

\section*{Acknowledgments}
SM is supported in part through the NExT Institute and STFC Consolidated Grant ST/X000583 /1. YH is supported by The Scientific and Technological Research Council of Turkey (TUBITAK) in the framework of the 2219-International Postdoctoral Research Fellowship Programme.

\appendix

\section{Detector performance for COHERENT data}
\label{app_coherentdata}

In this appendix, based on the original publications~\cite{COHERENT:2020iec,COHERENT:2020ybo,COHERENT:2021xmm,COHERENT:2023aln}, we give a short summary of the assumptions regarding the detector performance that has been made when analysing the COHERENT data. 
\subsection{Argon}
We use $N_{\rm POT}=1.38\times10^{23}$, $r=0.09$, L=27.5 m, $N=21.95$, $Z=18$, and $m_{\rm det}=24.4$ kg. The QF is given by 
\begin{equation}
{\rm QF}_{\rm Ar} = 0.24 + 0.00078 \, y \,E_{\max}^{\rm Ar}
\end{equation}
where $E_{\max}^{\rm Ar}=150$ is given in keV. The fractional ionisation energy is thus given by,
\begin{equation}
 y_{\rm ion}^{\rm Ar} = {\rm QF}_{\rm Ar} y \, .
\end{equation}

The energy resolution is in turn given by
\begin{equation}
R_{\rm Ar}(y_{\rm rec}^{\rm Ar},y_{\rm ion}^{\rm Ar})=\dfrac{2}{1+ {\rm Erf }(\frac{y_{\rm ion}^{\rm Ar} }{\sqrt{2} \sigma_{\rm Ar}})} \dfrac{1}{\sqrt{2\pi} \sigma_{\rm Ar} } \exp\left[ - \dfrac{(y_{\rm rec}^{\rm Ar} - y_{\rm ion}^{\rm Ar})^2}{2\sigma_{\rm Ar}^2 }\right],
\end{equation}
with the width $\sigma_{\rm Ar}$ given by
\begin{equation}
\sigma_{\rm Ar} = \dfrac{0.58y_{\rm ion}^{\rm Ar}}{\sqrt{( y_{\rm ion}^{\rm Ar} E_{\max}^{\rm Ar})}}
\end{equation}
where again $E_{\max}^{\rm Ar}=150$ is given in keV. 

Finally, the detector efficiency is parametrised by
\begin{equation}
 A_{\rm eff}^{\rm Ar}=0.96 \times \left(1 - \exp[a_{\rm Ar} + b_{\rm Ar} y_{\rm rec} ^{\rm Ar}+ c_{\rm Ar} (y_{\rm rec}^{\rm Ar})^2 + d_{\rm Ar} (y_{\rm rec}^{\rm Ar})^3]\right)
\end{equation}
where $a_{\rm Ar} = 0.336174$, $b_{\rm Ar} = -48.2704$, $c_{\rm Ar} = 301.634$, $d_{\rm Ar} = -4155.7$.

As a consistency check we have verified that integrating the nuclear recoil spectrum in the range $1.1$ keV$_{\rm ee}$ (which is where the efficiency goes to zero) to  $120.0$ keV$_{\rm ee}$ we get the expected number of recoils in the standard model with the COHERENT set up to be 132 which is in good agreement with the value $128\pm17$ given by the COHERENT collaboration. For comparison, the observed number of \cevns\ events by COHERENT is $159\pm43$.

\subsection{Caesium Iodine}

Here we use $N_{\rm POT}=3.2\times10^{23}$, $r=0.0848$, L=19.3 m, $N=76$, $Z=54$, and $m_{\rm det}=14.6$ kg. 
The QF is given by
\begin{equation}
{\rm QF}_{\rm CsI} = 0.0554628 + 4.30681( y E_{\max}^{\rm CsI})  - 111.707( y E_{\max}^{\rm CsI}) ^2 + 840.384( y E_{\max}^{\rm CsI}) ^3
\end{equation}
where $E_{\max}^{\rm CsI}=0.04642 $ is given in MeV. The fractional ionisation energy, then becomes
\begin{equation}
 y_{\rm ion}^{\rm CsI} = {\rm QF}_{\rm CsI} y
\end{equation}
and the energy resolution is given by
\begin{equation}
R_{\rm CsI}( {\rm PE},y_{\rm ion}^{\rm CsI})=
\dfrac{(a_{\rm CsI} (1 + b_{\rm CsI}))^{(1 + b_{\rm CsI})}}{\Gamma(1 + b_{\rm CsI})}  ({\rm PE})^{b_{\rm CsI}} \exp[-a_{\rm CsI} (1 + b_{\rm CsI}) {\rm PE}]
\end{equation}
where PE is the number of photo electrons, $a_{\rm CsI} = 0.0749/( y_{\rm ion}^{\rm CsI} E_{\max}^{\rm CsI}) = 0.00161/ y_{\rm ion}^{\rm CsI} $ and
$b_{\rm CsI}= 9.56( y_{\rm ion}^{\rm CsI} E_{\max}^{\rm CsI})=444 y_{\rm ion}^{\rm CsI} $  using $E_{\max}^{\rm CsI}=46.42$ in keV.

The detector efficiency is parametrised by
\begin{equation}
 A_{\rm eff}^{\rm CsI} = \left(\dfrac{a_{\rm CsI}^{\rm PE}}{1 + \exp[-b_{\rm CsI} ^{\rm PE}({\rm PE} -c_{\rm CsI}^{\rm PE})]} +  d_{\rm CsI}^{\rm PE} \right)\dfrac{a_{\rm CsI}^{\rm t} + \int_{a_{\rm CsI}^{\rm t}}^{c_{\rm CsI}^{\rm t}} \exp[-b_{\rm CsI} ^{\rm t}(t - a_{\rm CsI}^{\rm t})] dt }{c_{\rm CsI}^{\rm t}} 
\end{equation}
where $a_{\rm CsI}^{\rm PE} =1.32045$, $b_{\rm CsI} ^{\rm PE}= 0.285979$, $c_{\rm CsI}^{\rm PE} =  10.8646$, and $d_{\rm CsI}^{\rm PE} = -  0.33332$ whereas  $a_{\rm CsI}^{\rm t} =0.52 \, \mu$s, $b_{\rm CsI} ^{\rm t}= 0.0494 \, (\mu {\rm s})^{-1}$, and   $c_{\rm CsI}^{\rm t} =6.0 \, \mu$s. Thus, for simplicity we integrate over all times.

Again as a consistency check, integrating the nuclear recoil spectrum in the range $8$ $ {\rm PE}$ (which is essentially where the efficiency goes to zero) to  $60$ ${\rm PE}$ we get the expected number of recoils in the standard model with the COHERENT set up to be 335 which is in good agreement with the value $341\pm11\pm42$ given by COHERENT. For comparison, the observed number of \cevns\ events by COHERENT is $306\pm20$.

\subsection{Germanium}

When analysing the germanium data we use $N_{\rm POT} \, m_{\rm det} =2.09\times10^{23}$ kg, $r=0.288/3$, L=19.2~m, $N=40.6$, and $Z=32$. For the QF, the Lindhard model,  with $k=0.157$, is used in the same way as for the reactor data. For the intrinsic resolution of the detector we use  $\sigma_n({\rm SNS} )= 50.1$~eV$_{\rm ee}$ and the resolution is given by Eq.~(\ref{eq_ge_resolution}). 

Integrating the nuclear recoil spectrum in the range $1.5$ keV$_{\rm ee}$ to  $8.5$ keV$_{\rm ee}$ and $11.0$ keV$_{\rm ee}$ to  $20.0$ keV$_{\rm ee}$ we get the expected number of recoils in the standard model with the COHERENT set up to be 34.4 which is in good agreement with the value 35.1 given by the collaboration. The observed number of \cevns\ events by COHERENT is $20.6^{+7.1}_{-6.3}$.

\bibliographystyle{apsrev}  
\bibliography{refs} 

\end{document}